\documentclass[12pt]{article}
\begin{document}
\begin{flushleft}
{\it Mehta Research Institute}
\begin{flushright}
MRI-PHY/p970203\\
hep-th/9702170
\end{flushright}
\end{flushleft}
\begin{center}
\vspace*{1.0cm}

{\LARGE{\bf  $S$-matrices of non-simply laced affine Toda theories 
by folding}}

\vskip 1.5cm

{\large {\bf S. Pratik  Khastgir\footnote{email: pratik@mri.ernet.in}}}

\vskip 0.5cm

{\sl Mehta Research Institute for Mathematics and Mathematical Physics,} \\
{\sl Chhatnag Road, Jhunsi, Allahabad 221506, INDIA.}

\end{center}

\vspace{1 cm}

\begin{abstract}
\noindent 
The exact factorisable quantum $S$-matrices are known for simply laced 
as well as non-simply laced affine Toda field theories. 
Non-simply laced theories are obtained from the affine Toda theories
based on simply laced algebras by folding the corresponding Dynkin
diagrams. The same process, called classical `reduction', provides 
solutions of a non-simply laced theory from the classical solutions 
with special symmetries of the parent simply laced theory.
In the present note we shall elevate the idea
of folding and classical reduction to the quantum level. To support our 
views we have made some interesting observations for $S$-matrices of 
non-simply laced theories and give prescription for obtaining them 
through the folding of simply laced ones.

\end{abstract}

\vspace{1 cm}
\renewcommand{\theequation}{\arabic{section}.\arabic{equation}}
\def\ts{\thinspace}
\newcommand\et{{\sl et al. }}
\newcommand{\mbold}{\mbox{\boldmath$m$}}
\newcommand\beq{\begin{equation}}
\newcommand\eeq{\end{equation}}
\newcommand\Bear{\begin{eqnarray}}
\newcommand\Enar{\end{eqnarray}}
\newcommand{\rref}[1]{(\ref{#1})}
\def\hcd#1{\{ #1 \}'{}}
\newcommand\Zam{Zamolodchikov}
\newcommand \ZZ {A. B. \Zam\  and Al. B. \Zam, {\it Ann. Phys.}
{\bf 120} (1979)  253}
\newcommand\CMP{{\it Comm.\ts Math.\ts Phys.\ts}}
\newcommand\IJMP{{\it Int.\ts J.\ts Mod.\ts Phys.\ts}}
\newcommand\NP{{\it Nucl.\ts Phys.\ts}}
\newcommand\PL{{\it Phys.\ts Lett.\ts}}
\newcommand\Zm{Zamolodchikov}
\newcommand\AZm{A.\ts B.\ts \Zm}
\newcommand\dur{H.\ts W.\ts Braden, E.\ts Corrigan, P.\ts E.\ts Dorey 
and R.\ts Sasaki}


\font\dynkfont=cmsy10 scaled\magstep4    \skewchar\dynkfont='60
\def\dynk{\textfont2=\dynkfont}
\def\hr#1,#2;{\dimen0=.4pt\advance\dimen0by-#2pt
              \vrule width#1pt height#2pt depth\dimen0}
\def\vr#1,#2;{\vrule height#1pt depth#2pt}
\def\blb#1#2#3#4#5
            {\hbox{\ifnum#2=0\hskip11.5pt
           \else\ifnum#2=1\hr13,5.4;\hskip-1.5pt
           \else\ifnum#2=2\hr13.5,7.6;\hskip-13.5pt
                        \hr13.5,3.2;\hskip-2pt
                   \else\ifnum#2=3\hr13.7,8.4;\hskip-13.7pt
                                  \hr13,5.4;\hskip-13pt
                                  \hr13.7,2.4;\hskip-2.2pt
                   \else\ifnum#2=7\hr13.7,8.4;\hskip-13.7pt
                                  \hr13.2,6.4;\hskip-13.2pt
                                  \hr13.2,4.4;\hskip-13.2pt
                                  \hr13.7,2.4;\hskip-2.2pt
                   \else\ifnum#2=8\hr13.7,8.4;\hskip-13.7pt
                                  \hr13.2,6.4;\hskip-13.2pt
                                  \hr13.2,4.4;\hskip-13.2pt
                 \hr13.7,2.4;\hskip-8.2pt$\rangle$\hskip-2.0pt
                   \else\ifnum#2=4\hr13.5,7.6;\hskip-13.5pt
                \hr13.5,3.2;\hskip-8pt$\rangle$\hskip-1.8pt
                   \else\ifnum#2=5\hr30,5.4;\hskip-1.5pt         
                   \else\ifnum#2=6\hr13.7,8.4;\hskip-13.7pt
                                  \hr13,5.4;\hskip-13pt
               \hr13.7,2.4;\hskip-8.2pt$\rangle$\hskip-2.0pt
                      \fi\fi\fi\fi\fi\fi\fi\fi\fi
                   $#1$
                   \ifnum#4=0
                   \else\ifnum#4=1\hskip-9.2pt\vr22,-9;\hskip8.8pt
                   \else\ifnum#4=8\hskip-.4pt
                   \else\ifnum#4=2\hskip-10.9pt\vr22,-8.75;\hskip3pt
                                  \vr22,-8.75;\hskip7.1pt
                   \else\ifnum#4=3\hskip-12.6pt\vr22,-8.5;\hskip3pt
                                  \vr22,-9;\hskip3pt
                                  \vr22,-8.5;\hskip5.4pt
                   \else\ifnum#4=5\hskip-9.2pt\vr39,-9;\hskip8.8pt
                                  \fi\fi\fi\fi\fi\fi
                   \ifnum#5=0
                   \else\ifnum#5=1\hskip-9.2pt\vr1,12;\hskip8.8pt
                   \else\ifnum#5=8\hskip-9.2pt\vr1,25;\hskip8.8pt
                   \else\ifnum#5=2\hskip-10.9pt\vr1.25,12;\hskip3pt
                                  \vr1.25,12;\hskip7.1pt
                   \else\ifnum#5=3\hskip-12.6pt\vr1.5,12;\hskip3pt
                                  \vr1,12;\hskip3pt
                                  \vr1.5,12;\hskip5.4pt
                   \else\ifnum#5=5\hskip-9.2pt\vr1,29;\hskip8.8pt
                                   \fi\fi\fi\fi\fi\fi

                   \ifnum#3=0\hskip8pt
                   \else\ifnum#3=1\hskip-5pt\hr13,5.4;
                   \else\ifnum#3=2\hskip-5.5pt\hr13.5,7.6;
                                  \hskip-13.5pt\hr13.5,3.2;
                   \else\ifnum#3=3\hskip-5.7pt\hr13.7,8.4;
                                  \hskip-13pt\hr13,5.4;
                                  \hskip-13.7pt\hr13.7,2.4;
                  \else\ifnum#3=7\hskip-5.7pt\hr13.7,8.4;
                                 \hskip-13.2pt\hr13.2,6.4;
                                 \hskip-13.2pt\hr13.2,4.4;
                                  \hskip-13.7pt\hr13.7,2.4;
                  \else\ifnum#3=8\hskip-5.7pt\hr13.7,8.4;
                                 \hskip-13.2pt\hr13.2,6.4;
                                 \hskip-13.2pt\hr13.2,4.4;
                    \hskip-13.9pt$\langle$\hskip-8pt\hr13.7,2.4;
                   \else\ifnum#3=4\hskip-5.5pt\hr13.5,7.6;
                                  \hskip-13.5pt$\langle$\hskip-8.2pt
                                  \hr13.5,3.2;
                  \else\ifnum#3=5\hskip-5pt\hr30,5.4;
                   \else\ifnum#3=6\hskip-5.7pt\hr13.7,8.4;
                                  \hskip-13pt\hr13,5.4;
                       \hskip-13.9pt$\langle$\hskip-8.0pt\hr13.7,2.4;
                                 \fi\fi\fi\fi\fi\fi\fi\fi\fi
                   }}
\def\blob#1#2#3#4#5#6#7{\hbox
{$\displaystyle\mathop{\blb#1#2#3#4#5 }_{#6}\sp{#7}$}}
\def\up#1#2{\dimen1=33pt\multiply\dimen1by#1
                  \hbox{\raise\dimen1\rlap{#2}}}
\def\uph#1#2{\dimen1=17.5pt\multiply\dimen1by#1
                  \hbox{\raise\dimen1\rlap{#2}}}
\def\dn#1#2{\dimen1=33pt\multiply\dimen1by#1
                   \hbox{\lower\dimen1\rlap{#2}}}
\def\dnh#1#2{\dimen1=17.5pt\multiply\dimen1by#1
                    \hbox{\lower\dimen1\rlap{#2}}}

\def\rlbl#1{\kern-8pt\raise3pt\hbox{$\scriptstyle #1$}}
\def\clbl#1{\kern-8pt\hbox{$\scriptstyle #1$}}
\def\llbl#1{\raise3pt\llap{\hbox{$\scriptstyle #1$\kern-8pt}}}
\def\elbl#1{\kern3pt\lower4.5pt\hbox{$\scriptstyle #1$}}
\def\lelbl#1{\rlap{\hbox{\kern-9pt\raise2.5pt\hbox{{$\scriptstyle #1$}}}}}
\def\elblr#1{\rlap{\hbox{\kern-18pt\lower5pt\hbox{{$\scriptstyle #1$}}}}}
\def\elblc#1{\rlap{\hbox{\kern-15pt\hbox{{$\scriptstyle #1$}}}}}
\def\elbld#1{\rlap{\hbox{\kern-24pt\hbox{{$\scriptstyle #1$}}}}}

\def\wht#1#2#3#4{\blob\circ#1#2#3#4{}{}}
\def\whtd#1#2#3#4#5{\blob\circ#1#2#3#4{#5}{}}
\def\whtu#1#2#3#4#5{\blob\circ#1#2#3#4{}{#5}}
\def\whtr#1#2#3#4#5{\blob\circ#1#2#3#4{}{}\rlbl{#5}}
\def\whtc#1#2#3#4#5{\blob\circ#1#2#3#4{}{\clbl{#5}}}
\def\whtl#1#2#3#4#5{\llbl{#5}\blob\circ#1#2#3#4{}{}}

\def\drwng#1#2#3{\hbox{$\vcenter{ \offinterlineskip{
  \hbox{\phantom{}\kern6pt{$\circ^{\elbl{#3}}$}}
  \kern-2.5pt\hbox{$\Biggr/$}\kern-0.5pt
  \hbox{\phantom{}\kern-5pt$\circ^{ \elbl{#1}}$}
  \kern-3.0pt\hbox{$\Biggr\backslash$}
  \kern-1.5pt\hbox{\phantom{}\kern6pt{$\circ^{\elbl{#2}}$}}  } }$}}

\def\drwngl#1#2#3{\hbox{$\vcenter{ \offinterlineskip{
  \hbox{\phantom{}\kern6pt{$\circ^{\elblr{#3}}$}}
  \kern-2.5pt\hbox{$\Biggr/$}\kern-3.5pt
  \hbox{\phantom{}\kern-5pt$\circ^{\elblc{#1}}$}
  \kern-3.0pt\hbox{$\Biggr\backslash$}
  \kern-1.5pt\hbox{\phantom{}\kern6pt{$\circ^{\elblr{#2}}$}}  } }$}}

\def\drwngd#1#2#3{\hbox{$\vcenter{ \offinterlineskip{
  \hbox{\phantom{}\kern6pt{$\circ^{\elblr{#3}}$}}
  \kern-2.5pt\hbox{$\Biggr/$}\kern-3.5pt
  \hbox{\phantom{}\kern-5pt$\circ^{\elbld{#1}}$}
  \kern-3.0pt\hbox{$\Biggr\backslash$}
  \kern-1.5pt\hbox{\phantom{}\kern6pt{$\circ^{\elblr{#2}}$}}  } }$}}

\def\rde#1#2#3{\raisebox{.5pt}{\hbox{\phantom{}\kern-4pt\hbox{$\vcenter
{\offinterlineskip\hbox{
               \raise 4.5pt\hbox{\vrule height0.4pt width13pt depth0pt}
                \kern-1pt\vbox{ \hbox{\drwng{#1}{#2}{#3}}} }}$  }} }}

\def\rdel#1#2#3{\raisebox{.5pt}{\hbox{\phantom{}\kern-4pt\hbox{$\vcenter
{\offinterlineskip\hbox{
               \raise 4.5pt\hbox{\vrule height0.4pt width13pt depth0pt}
                \kern-1pt\vbox{ \hbox{\drwngl{#1}{#2}{#3}}} }}$  }} }}

\def\rded#1#2#3{\raisebox{.5pt}{\hbox{\phantom{}\kern-4pt\hbox{$\vcenter
{\offinterlineskip\hbox{
               \raise 4.5pt\hbox{\vrule height0.4pt width13pt depth0pt}
                \kern-1pt\vbox{ \hbox{\drwngd{#1}{#2}{#3}}} }}$  }} }}

\def\ddgiid#1.#2.{\dynk  \whtd0300{#1}\whtd6000{#2}}

\def\ddgiiu#1.#2.{\dynk  \whtu0300{#1}\whtu6000{#2}}

\def\ddfid#1.#2.#3.#4.{\dynk\whtd0100{#1}%
               \whtd1200{#2}\whtd4100{#3}\whtd1000{#4}}

\def\ddfiu#1.#2.#3.#4.{\dynk\whtu0100{#1}%
               \whtu1200{#2}\whtu4100{#3}\whtu1000{#4}}

\def\ddands#1.#2.#3.#4.#5.{\dynk \whtd0100{#1}\whtd1100{#2}\hbox%
 {$\cdot\cdot$}\whtd1100{#3}\hbox{$\cdot\cdot$}\whtd1100{#4}\whtd1000{#5}}

\def\ddanus#1.#2.#3.#4.#5.{\dynk \whtu0100{#1}\whtu1100{#2}\hbox%
 {$\cdot\cdot$}\whtu1100{#3}\hbox{$\cdot\cdot$}\whtu1100{#4}\whtu1000{#5}}

\def\ddbnidr#1.#2.#3.#4.#5.{\dynk \whtd0100{#1}\whtd1100{#2}%
             \hbox{$\cdot\cdot$}\whtd1100{#3}\whtd1200{#4}\whtd4000{#5}}

\def\ddbniur#1.#2.#3.#4.#5.{\dynk \whtu0100{#1}\whtu1100{#2}%
             \hbox{$\cdot\cdot$}\whtu1100{#3}\whtu1200{#4}\whtu4000{#5}}

\def\ddcnd#1.#2.#3.#4.#5.{\dynk \whtd0100{#1}\whtd1100{#2}%
    \hbox{$\cdot\cdot$}\whtd1100{#3}\whtd1400{#4}\whtd2000{#5}}

\def\ddcnu#1.#2.#3.#4.#5.{\dynk \whtu0100{#1}\whtu1100{#2}%
      \hbox{$\cdot\cdot$}\whtu1100{#3}\whtu1400{#4}\whtu2000{#5}}

\def\dddndt#1.#2.#3.#4.#5.#6.{\hbox{$\vcenter{\hbox
         {\dynk\hbox{$ \whtd0100{#1}\whtd1100{#2}\hbox{$\cdot\cdot$}%
          \whtd1100{#3}\rde{#4}{#5}{#6} $}} }$}}

\def\dddnut#1.#2.#3.#4.#5.#6.{\hbox{$\vcenter{\hbox%
         {\dynk\hbox{$ \whtu0100{#1}\whtu1100{#2}\hbox{$\cdot\cdot$}%
          \whtu1100{#3}\rded{#4}{#5}{#6} $}} }$\kern16.65pt}}

\def\dddf#1.#2.#3.#4.{\hbox{$\vcenter{\hbox
         {\dynk\hbox{$\whtd0100{#1}\rde{#2}{#3}{#4} $}} }$}}

\def\dduf#1.#2.#3.#4.{\hbox{$\vcenter{\hbox
         {\dynk\hbox{$\whtu0100{#1}\rdel{#2}{#3}{#4} $}} }$\kern16.65pt}}

\def\ddei#1.#2.#3.#4.#5.#6.{\hbox{$\vcenter{\hbox
       {\dynk \whtd0100{#1}\whtd1100{#3}%
       \up1{\whtr0001{#2}}\whtd1110{#4}\whtd1100{#5}\whtd1000{#6}} }$}}

\def\ddeiu#1.#2.#3.#4.#5.#6.{\raise.21pt\hbox{$\vcenter{\hbox
       {\dynk \whtu0100{#1}\whtu1100{#3}%
       \up1{\whtl0008{#2}}\whtc1180{#4}\whtu1100{#5}\whtu1000{#6}} }$%
      \kern.26pt}}

\baselineskip=18pt

\section{Introduction}
\setcounter{equation}{0}

Affine Toda field theories \cite{MOPa,AFZa} received considerable attention 
in past years. At present it is one of the best understood field theories 
at classical and quantum levels. 
The renewal of interest for Toda field theories in recent years is due 
to the work of Zamolodchikov \cite{Zam}, where he discussed a class of
deformation of conformal field theories that preserved integrabilty and 
further showed that resultant theory is characterised by eight masses 
related to the Cartan matrix of $E_8$, and by integrals of motion with 
spins given by the exponents of $E_8$ modulo its Coxeter number.
The natural description of these resultant integrable theories with 
massive excitations is in terms of their $S$-matrices. Subsequently, 
Hollowood and Mansfield \cite{HMBL} considered a class of integrable 
field theories, namely Toda field theories and showed that for 
particular values of the coupling constant these theories describe the
minimal models. The affine version of these theories was connected to 
the perturbed conformal field theory. This basically motivated people 
to study affine Toda field theories based on various Lie algebras
[5--11].
Toda field theory is integrable at the classical level \cite{MOPa,OTa}
due to the presence of an infinite number of conserved quantities. 
It is firmly believed
that the integrability survives quantisation. 
Higher-spin quantum conserved currents are discussed in
Ref. \cite{DGZa}. Exact quantum $S$-matrices for 
affine Toda field theories based on simply laced algebras, $a_n^{(1)}$,
$d_n^{(1)}$, $e_{6,7,8}^{(1)}$ were constructed successfully in Refs.
\cite{BCDSc}, but after that one had to wait more than two years for 
the $S$-matrices of non-simply laced theories, because of their intricate
nature. Delius \et \cite{DGZc} came up with the beautiful idea
of floating masses and constructed $S$-matrices for most of the non-simply 
laced theories. The 
remaining ones were constructed by Corrigan \et \cite{CDS} where 
generalised bootstrap principle is introduced and more insight
to the mechanism is provided. The singularity structure of the  
$S$-matrices of simply laced theories, which in some   
cases contain poles up to 12-th order \cite{BCDSc},
is  beautifully explained in terms of the
singularities of the corresponding Feynman diagrams
\cite{BCDSe}, so called Landau singularities.
 
Non-simply laced theories are obtained from the affine Toda theories
based on simply laced algebras by folding the corresponding Dynkin
diagrams. The same process, called classical `reduction', provides 
solutions of a non-simply laced theory from the classical solutions 
with special symmetries of the parent simply laced theory.
In the present paper we construct the $S$-matrices of non-simply laced
theories by folding the simply laced theories. For this we have to 
write the $S$-matrices of the simply
laced ones cleverly and just substitute the Coxeter number of non simply
laced theories for obtaining the S-matrix. This is a new type of 
construction, hint of which was already present in Ref.\cite{BCDSc},
but authors left them at very early stage. We think that this kind of 
construction will provide better understanding of the structure of 
$S$-matrices and mechanism. This paper is organised as follows. 
In the next section we give brief
introduction of Toda theories and the associated $S$-matrices. 
Section 3 contains our new formulae for $S$-matrices incorporating
the idea of folding. In the following section we consturct the
$S$-matrices of a specific example of $c_n^{(1)}$ theory.
In subsequent sections we explicitly show the way of constructing 
$S$-matrices for $f_4^{(1)}$, $b_n^{(1)}$ and $g_2^{(1)}$ theories 
through folding. In section 8 we discuss twisted theories and 
final section is reserved for some discussions.


\section{Affine Toda field theory: An overview}

In this section we give a quick review of Toda field
theories.\footnote{For an excellent review see Ref. \cite{Co}.}
Affine Toda field theory \cite{MOPa} is a massive scalar field theory with 
exponential interactions in $1+1$ dimensions described by the Lagrangian
\begin{equation}
{\cal L}={1\over 2}
\partial_\mu\phi^a\partial^\mu\phi^a-{m^2\over
\beta^2}\sum_{i=0}^rn_ie^{\beta\alpha_i\cdot\phi}.
\label{ltoda}
\end{equation}
The field $\phi$ is an $r$-component scalar field, $r$ is the rank of a
compact semi-simple Lie algebra $g$ with $\alpha_i$;
$i=1,\ldots,r$ being its simple roots and $\alpha_0$ is the affine root. 
The roots are normalised so that long roots have length 2, $\alpha_L^2=2$.
The Kac-Coxeter labels $n_i$ are such that $\sum_{i=0}^rn_i\alpha_i=0$, 
with the convention $n_0=1$. The quantity, $\sum_{i=0}^rn_i$, 
is denoted by `$h$' and known as the Coxeter number.
When the term containing the affine root is removed, the theory becomes
conformally invariant (conformal Toda field theory). Then the theory
is based on the root system of a finite Lie algebra `$g$' and sometimes
it is called a non-affine Toda theory in distinction with the affine one.
`$m$' is a real parameter setting the mass scale of the theory 
and $\beta$ is a real coupling constant,
which is relevant only in quantum theory.
The equation of motion reads, 
\beq
 \partial\sp2\phi = -{\frac{m^2}{\beta}}\sum_{i=0}\sp r n_i\alpha_ie\sp
{\beta\alpha_i\cdot\phi} . \label{eom}
\eeq
It turns out that the data in quantum theory, 
such as the masses and couplings
of various kinds, are also useful for the reduction of
classical equation of motion.
Expanding the potential part  
of the Lagrangian up to second order, one can extract a 
$({\rm mass})\sp2$ matrix 
\beq
(M\sp2)\sp{ab}=m^2\sum\sp r_{i=0}n_i\alpha_i\sp a\alpha_i\sp b . 
\label{mass}\eeq
The mass matrix has been studied before \cite{MOPa,BCDSc,CMa,KMb,BCDSb}.
One important fact which underlies the present work is that
the particles of the simply laced theory 
are associated unambiguously with the spots on the Dynkin diagram 
and thus to the simple roots (fundamental weights) of 
the associated finite Lie algebra [6, 17--19].
It is based on the
observation that the set of masses computed
as the $r$ eigenvalues of the mass matrix \rref{mass} actually
constitute the Frobenius-Perron eigenvector of the Cartan matrix of 
the associated finite Lie algebra. In other words, if we set
${\mbold}=(m_1,m_2,\dots ,m_r)$
then
\beq 
C{\mbold}=\lambda_{\rm min}{\mbold}=4m^2\sin\sp2{\pi\over 2h}\ {\mbold}, 
\label{fp}
\eeq
where $C$ is the Cartan matrix
$C_{ij}=2\alpha_i\cdot\alpha_j/\alpha_j\sp2$,  
$i,j=1,\dots ,r. $
The Coxeter numbers and $({\rm mass})\sp2$ of various theories
together with the Dynkin diagrams and particle labelling, 
can be found in Ref. \cite{BCDSc}. 

Folding and reductions based on the symmetry 
(automorphism) of the Dynkin diagram can be understood in the
following way \cite{OT,BCDSc}:
a symmetry of the Dynkin diagram, permuting the
points as $\alpha\rightarrow p(\alpha)$, can be rewritten as a mapping 
of the field space to itself, $\phi\rightarrow p(\phi)$. This is a 
symmetry of the classical field equations \rref{eom}, namely it 
maps a solution to another. 
This means that if the fields initially take 
values in the subspace invariant under $p$, they will remain there, 
at least classically. Since the subspace is of smaller dimension than 
the original field space, the evolution of fields within it can be 
described in terms of an equation with fewer variables than the
original equation. The latter
is obtained by projecting the variables $\alpha_i$ in eqn. \rref{eom} 
onto the invariant subspace. This process of obtaining new equations 
and their solutions from the old, by exploiting diagram symmetries, 
is known as reduction.
In other words, an arbitrary solution of a reduced theory always gives 
solution(s) of the original theory by appropriate embedding(s).
The so-called direct reductions are those such that $\alpha\cdot p(\alpha
)=0$ for each root $\alpha$ (i.e. the symmetry does not relate
points linked by a line on the Dynkin diagram).
A symmetry of the unextended Dynkin diagram of a
simply-laced algebra yields the diagram for one of the non simply-laced
algebras on projection onto the invariant subspace, and the addition of
the extra point to extend the diagram always respects such a symmetry. 
The resulting projected diagram is the untwisted affine diagram for 
the non simply-laced algebra. This is a reflection of the fact 
that the symmetry group of the extended diagram contains always at 
least that of the unextended diagram. Reductions involving any
additional symmetries of the extended diagram yield affine Toda theories
based on the twisted affine Dynkin diagrams. The reductions based on 
the symmetries of the unextended as well as extended (affine) Dynkin 
diagrams are discussed rather completely in earlier papers [6, 20--22].
There is an interesting distinction to be made
here between the two different types of non simply-laced theories, 
namely twisted and untwisted theories. 
The foldings leading to untwisted theories turn out to remove 
degeneracies from the mass spectrum, the resulting non degenerate
particles always being linear combinations of the degenerate particles 
in the parent theory. In contrast, foldings leading to twisted diagrams 
remove some particles from the spectrum altogether, while leaving 
the others unchanged.

Quantum $S$-matrices of all affine Toda theories are also known
[2, 5--9].
 Based on the assumption that the infinite set of conserved quantities
be preserved after quantisation, only the elastic processes are
allowed and the multi-particle $S$-matrices are factorised into 
a product of two particle elastic $S$-matrices.
A typical elastic, unitary $S$-matrix for a process $a+b\rightarrow a+b$
can be written as product of ratios of hyperbolic sines.
\beq
S_{ab}(\theta)=\prod_{x\in I_{ab}} \{x\}, \qquad 
\{x\}={{(x-1)(x+1)}\over{(x-1+B)(x+1-B)}},
\label{smat}
\eeq         
for some set of integers $I_{ab}$. Block $(x)$ and the function 
$B(\beta)$ are given by
\beq
(x)={{\sinh({\frac{\theta}{2}+\frac{i\pi}{2h}x)}}\over
{\sinh({\frac{\theta}{2}-\frac{i\pi}{2h}x)}}},\qquad
B(\beta)={{\frac{1}{2\pi}}\frac{\beta^2}{1+\beta^2/4\pi}}.
\label{block}
\eeq
$\theta=\theta_a-\theta_b$ is the relative rapidity
($p_a\equiv(m_a\cosh{\theta_a},m_a\sinh{\theta_a})$), $h$ is the Coxeter 
number of the Lie algebra on which theory is based.
For $x\leq h$, mod $2h$ these $S$-matrices have physical sheet simple poles
at $\theta={{i\pi x}/{h}}$ and these can be interpreted as elementary 
particle poles from $s$-channel or $u$-channel exchange, with mass 
related to the value of $x$. Above $S$-matrices respect crossing symmetry
and bootstrap principle \cite{BCDSc}. 
$S$-matrices for various simply laced theories were constructed in Refs.
[6--9].
But the ideas of simply laced theories failed for the theories based
on non-simply laced algebras. The problem was resolved by Delius \et
\cite{DGZc} with the introduction of floating masses and the 
renormalised Coxeter number. Next the `generalised bootstrap principle'
was introduced in Ref. \cite{CDS} and construction of all the
$S$-matrices for various non-simply laced theories was completed.
According to the `generalised bootstrap principle' there is a quantum
field theory corresponding to the dual pair of non-simply laced
algebras together than either classical theory separately and the 
weak ($\beta\rightarrow 0$) and strong ($\beta\rightarrow \infty$)
coupling limits of this quantum field theory would effectively
lead to one or the other of the dual pair. In other
words the transformation $\beta\rightarrow 4\pi/\beta$ effectively
implements the inversion $\alpha_i\rightarrow 2\alpha_i/|\alpha_i|^2$,
which interchanges the two extended Dynkin diagrams. For every dual
pair one defines a renormalised Coxeter number $H$, which is a 
function of $B(\beta)$ and interpolates the Coxeter numbers of
the dual algebras.
In what follow we shall be following the notations of Ref. \cite{BCDSc},
which is convenient. We shall be mentioning the explicit forms of 
$S$-matrices for various types of theories as we proceed.
For future convenience we arm ourselves with the following notations,
\beq
(x)_H={{\sinh({\frac{\theta}{2}+\frac{i\pi}{2H}x)}}\over
{\sinh({\frac{\theta}{2}-\frac{i\pi}{2H}x)}}},\qquad
\{x\}_{\nu}={{(x-\nu B-1)_H(x+\nu B+1)_H}\over
{(x+\nu B+B-1)_H(x-\nu B-B+1)_H}}
\label{bloch}
\eeq
and
\beq
[x]_{\nu}=\{x\}_{\nu}\{H-x\}_{\nu}, 
\label{bloch1}
\eeq
where $H$ is the renormalised Coxeter number described in 
Refs. \cite{DGZc,CDS}.

\section {$S$-matrices}
\setcounter{equation}{0}
First we consider non-simply laced algebras which are not twisted, (that is 
Dynkin diagrams for these non-simply laced algebras can be obtained  
by folding the non-affine simply laced algebras using the automorphisms of
Dynkin diagrams). In Ref. \cite{BCDSc} it was argued that 
$S$-matrices of the corresponding affine theories might be related by 
following formula.
\beq
S_{ab}^{\rm daughter}= S_{ab}^{\rm parent} S_{a{\bar b}}^{\rm parent}.
\label{datpran} 
\eeq
But this formula could not explain multipole structure on the basis of 
daughter Lagrangian.                 
Here we propose a slight modification to the formula \rref{datpran}, we
would only replace in the right hand side of \rref{datpran} the Coxeter 
number $h$, of the parent theory by the Coxeter number $H$ of the daughter 
theory. This new $H$ is a kind of renormalized Coxeter number as 
suggested by Delius \et in their paper \cite{DGZc}.
So, we have,
\beq
S_{ab}^{\rm daughter}= S_{ab}^{\rm parent} S_{a{\bar b}}^{\rm parent}
|_{h\rightarrow H}.
\label{datpram} 
\eeq
But to show that the formula is universal first one has to write down the
$S$-matrices of the parent theory cleverly in terms of Coxeter number
$h$. Terms written in arbritary fashion would lead to wrong results. 
Although we don't have any general rule for writing them yet, in the 
following we will show (case by case that) that the above formula
holds good. Furthermore we would see that although the formula 
\rref{datpram} is nice for checking the bootstrap, but it conceals a 
number of cancellations between zeros and poles (see Ref.\cite{CDS}). 
Finally we propose a set formulae, which makes more sense physically, 
by modifying the formula \rref{datpram} a little further. That is
\beq
S_{ab}^{\rm daughter}= S_{ab}^{\rm parent} S_{a{\bar b}}^{\rm parent}
|_{h\rightarrow H},\nonumber 
\label{nfora} \eeq
\beq
S_{an}^{\rm daughter}= S_{an}^{\rm parent}|_{h\rightarrow H},\nonumber
\label{nforb} \eeq
\beq
S_{mn}^{\rm daughter}= S_{mn}^{\rm parent}|_{h\rightarrow H}.
\label{nfor} 
\eeq
Above $a,b$ represent degenerate particles of the parent 
theory (non-degenerate in daughter) and $m,n$ are non-degenerate 
particles of the parent theory. For example let us consider the folding
of $e_6^{(1)}$ into $f_4^{(1)}$,\footnote{ Although we talk about folding of 
affine algebras, the figures we present will depict the folding of 
non-affine Dynkin diagrams only. The folding of corresponding affine
diagrams is just adding the affine vertex at proper positions \cite{BCDSc}.
The purpose of giving these diagrams is to provide mass labellings.}

\beq
\ddei{m_1}.{m_2}.{m_3}.{m_4}.{m_3}.{m_1}. 
\llap {\ddeiu{1}.{2}.{3}.{4}.{\bar 3}.{\bar 1}.}
\Rightarrow \ddfid {m_2}.{m_4}.{m_3}.{m_1}.
\llap{\ddfiu {2}.{4}.{3}.{1}.}
\label{ef}
\eeq

In the above case $a,b=\{1,3\}$ and  $m,n=\{2,4\}$.
At this point let us compare the expression \rref{datpram} with expressions
\rref{nfora}--\rref{nfor}.
The folding, leading to untwisted theories turns out to remove degeneracies
from the mass spectrum, the resulting non-degenerate particles always 
being linear combinations of the degenerate particles of the parent theory.
Since all the particles in the daughter theory are self conjugate,  
expression \rref{datpram} is manifestly crossing symmetric. This can be
seen in the following way. We know 
\beq
S_{a\bar b}(\theta)=S_{ab}(i\pi-\theta),
\eeq
at the level of the building block $\{x\}$, it is realised by,
\beq
\{x\}_{i\pi-\theta}=\{h-x\}_{\theta}.
\eeq
As mentioned in earlier section (expression \rref{smat}), $S_{ab}$ is
a product of these building blocks, therefore the combination
$S_{ab}^{\rm parent}S_{a\bar b}^{\rm parent}$, appearing in \rref{datpram}
always contains $\{x\}$ in the self conjugate combination:
\beq
\{x\}\{h-x\}~(\equiv x).
\eeq
Now coming to the formulae \rref{nfora}--\rref{nfor}. For expression 
\rref{nfora} arguments are the same as above and it is manifestly crossing 
symmetric because $a,~b$ belong to degenerate class of particles in 
the parent theory and become a self conjugate and non-degenerate in the 
daughter theory. On the other hand $m,~n$ were non-degenerate and self 
conjugate in the parent itself and therefore 
$S_{mn}^{\rm parent}$ must have 
been crossing symmetric to start with. Similarly,
$S_{an}^{\rm parent}(\theta)=S_{an}^{\rm parent}(i\pi-\theta)$ 
(because $n$ is self conjugate particle in parent theory), 
so $S_{an}^{\rm parent}$ is also
crossing symmetric. That is why \rref{nforb} and \rref{nfor} have only
one term, whereas \rref{nfora} has two terms on the right hand side
respectively and as expected all the $S$-matrices of the daughter theory
are crossing symmetric.

Twisted algebras are obtained by folding the affine versions of simply
laced algebras. These algebras use the extra symmetry of the affine 
Dynkin diagrams (automorphisms). In contrast to untwisted theories,
folding leading to twisted diagrams remove some particles from the
spectrum altogether \cite{BCDSc}, while leaving the others unchanged.
The particles surviving in the daughter form a subset of the parent 
theory mass spectrum.
So in this case \rref{nfor} is applicable and resulting $S$-matrices
form a subset of the original $S$-matrices of the the parent theory
with the Coxeter $h$ replaced by the renormalised Coxeter $H$.
\section {$a_{2n-1}^{(1)}\rightarrow c_n^{(1)}$}
\setcounter{equation}{0}

First we take the case of $c_n^{(1)}$ theories. 
The Dynkin diagram of $c_n^{(1)}$
is obtained by folding $a_{2n-1}^{(1)}$. Mass labelling and folding are
shown in following diagram.

\beq
\ddands{m_1}.{m_2}.{m_n}.{m_2}.{m_1}. 
\llap{\ddanus {1}.2.n.{\bar 2}.{\bar 1}.} 
\Rightarrow
\ddcnd{m_1}.{m_2}.{m_{n-2}}.{m_{n-1}}.{m_n}.
\llap{\ddcnu {1}.{2}.{n-2}.{n-1}.{n}.}
\label{atc}
\eeq

To be more specific let us take example of 
$a_5^{(1)}\rightarrow c_3^{(1)}$. In $a_5^{(1)}$ there are $5$ particles, 
$1^{\rm st}$ and $2^{\rm nd}$ particles
come with their mass degenerate conjugate partners, 
$5^{\rm th}$(or $\bar 1$)
and $4^{\rm th}$(or $\bar 2$) respectively.
The $3^{\rm rd}$ particle is self conjugate and the Coxeter number $h=6$. 
In $c_3^{(1)}$ the 
mass degeneracy is removed
and one has three self conjugate particles.The 
renormalized Coxeter $H=6+B$ for the dual pair ($c_3^{(1)}$, $d_4^{(2)}$).
Now we write the $S$-matrices of $c_3^{(1)}$ in the following manner. 
\beq
S_{11}^{c_3^{(1)}}=S_{11}^{a_5^{(1)}}S_{1\bar 1}^{a_5^{(1)}}
|_{h\rightarrow H}
=\{1\}\{h-1\}|_{h\rightarrow H}=\{1\}_0\{H-1\}_0
=[1]_0.
\eeq
Above is the correct answer for $S_{11}^{c_3^{(1)}}$ (see
Ref.\cite{CDS}). Similarly, 

\Bear
&S_{12}^{c_3^{(1)}} &= S_{12}^{a_5^{(1)}}S_{1\bar 2}^{a_5^{(1)}}
|_{h\rightarrow H}
=\{2\}\{h-2\}|_{h\rightarrow H}=\{2\}_0\{H-2\}_0=[2]_0,\nonumber\\
&S_{22}^{c_3^{(1)}}&= S_{22}^{a_5^{(1)}}S_{2\bar 2}^{a_5^{(1)}}
|_{h\rightarrow H}
=\{1\}\{3\}\{h-1\}\{h-3\}|_{h\rightarrow H}\nonumber\\
&&=\{1\}_0\{3\}_0\{H-1\}_0\{H-3\}_0=[1]_0[3]_0,\nonumber\\
&S_{31}^{c_3^{(1)}}& = S_{31}^{a_5^{(1)}}S_{3\bar 1}^{a_5^{(1)}}
|_{h\rightarrow H}
=\{3\}\{h-3\}|_{h\rightarrow H}=\{3\}_0\{H-3\}_0=[3]_0,\nonumber\\
&S_{32}^{c_3^{(1)}}& = S_{32}^{a_5^{(1)}}S_{3\bar 2}^{a_5^{(1)}}
|_{h\rightarrow H}
=\{2\}\{4\}\{h-2\}\{h-4\}|_{h\rightarrow H}\nonumber\\
&&=\{2\}_0\{4\}_0\{H-2\}_0\{H-4\}_0=[2]_0[4]_0,\nonumber\\
&S_{33}^{c_3^{(1)}}& = S_{33}^{a_5^{(1)}}S_{33}^{a_5^{(1)}}
|_{h\rightarrow H}
=\{1\}\{3\}\{5\}\{h-1\}\{h-3\}\{h-5\}|_{h\rightarrow H}\nonumber\\
&&=\{1\}_0\{3\}_0\{5\}_0\{H-1\}_0\{H-3\}_0\{H-5\}_0=[1]_0[3]_0[5]_0.
\Enar
Now we  show that the above  expressions can also be expressed as 
\rref{nfora}--\rref{nfor}.
The first three $S$-matrices remain the same because now $a,b=1,2$
and $m,n=3$. In the last three cancellations take  place and they
can be written in the following way. For this one has to go to the
basic building blocks and write them cleverly to obtain the answer.

\Bear
&S_{31}^{c_3^{(1)}}& = S_{31}^{a_5^{(1)}}|_{h\rightarrow H}
=\{3\}|_{h\rightarrow H}
={\frac{(2)(4)}{(2+B)(4-B)}}{\Big |}_{h\rightarrow H}
={\frac{(2)(h-2)}{(h-4+B)(4-B)}}{\Big |}_{h\rightarrow H}\nonumber\\
&&={\frac{(2)_H(H-2)_H}{(H-4+B)_H(4-B)_H}}={\frac{(2)_H(4+B)_H}
{(2+2B)_H(4-B)_H}}\nonumber\\
&&={\frac{(2)_H(4)_H}{(2+B)_H(4-B)_H}}{\frac{(2+B)_H(4+B)_H}
{(2+2B)_H(4)_H}}\nonumber\\
&&={\frac{(2)_H(4)_H}{(2+B)_H(4-B)_H}}{\frac{(H-4)_H(H-2)_H}
{(H-4+B)_H(H-2-B)_H}}
=\{3\}_0\{H-3\}_0=[3]_0,\nonumber\\
&S_{32}^{c_3^{(1)}}& = S_{32}^{a_5^{(1)}}|_{h\rightarrow H}
=\{2\}\{4\}|_{h\rightarrow H}
={\frac{(1)(3)}{(1+B)(3-B)}}{\frac{(3)(5)}{(3+B)(5-B)}}
{\Big |}_{h\rightarrow H}
\nonumber\\
&&={\frac{(1)(h-3)}{(h-5+B)(3-B)}}{\frac{(3)(h-1)}{(h-3+B)(5-B)}}
{\Big |}_{h\rightarrow H}\nonumber\\
&&={\frac{(1)_H(H-3)_H}{(H-5+B)_H(3-B)_H}}{\frac{(3)_H(H-1)_H}
{(H-3+B)_H(5-B)_H}}\nonumber\\
&&={\frac{(1)_H(3+B)_H}{(1+2B)_H(3-B)_H}}{\frac{(3)_H(5+B)_H}
{(3+2B)_H(5-B)_H}}\nonumber\\
&&={\frac{(1)_H(3)_H}{(1+B)_H(3-B)_H}}{\frac{(3+B)_H(5+B)_H}
{(3+2B)_H(5)_H}}\nonumber\\
&&~\times{\frac{(3)_H(5)_H}{(3+B)_H(5-B)_H}}{\frac{(1+B)_H(3+B)_H}
{(1+2B)_H(3)_H}}\nonumber\\
&&=\{2\}_0\{4+B\}_0\{4\}_0\{2+B\}_0=\{2\}_0\{H-2\}_0\{4\}_0\{H-4\}_0
=[2]_0[4]_0,\nonumber\\
&S_{33}^{c_3^{(1)}}& = S_{33}^{a_5^{(1)}}|_{h\rightarrow H}
=\{1\}\{3\}\{5\}|_{h\rightarrow H}\nonumber\\
&&={\frac{(0)(2)}{(B)(2-B)}}{\frac{(2)(4)}{(2+B)(4-B)}}
{\frac{(4)(6)}{(4+B)(6-B)}}{\Big |}_{h\rightarrow H}\nonumber\\
&&={\frac{(0)(h-4)}{(h-6+B)(2-B)}}{\frac{(2)(h-2)}{(h-4+B)(4-B)}}
{\frac{(4)(h)}{(h-2+B)(6-B)}}{\Big |}_{h\rightarrow H}\nonumber\\
&&={\frac{(0)_H(H-4)_H}{(H-6+B)_H(2-B)_H}}{\frac{(2)_H(H-2)_H}
{(H-4+B)_H(4-B)_H}}
{\frac{(4)_H(H)_H}{(H-2+B)_H(6-B)_H}}\nonumber\\
&&={\frac{(0)_H(2+B)_H}{(2B)_H(2-B)_H}}
{\frac{(2)_H(4+B)_H}{(2+2B)_H(4-B)_H}}
{\frac{(4)_H(6+B)_H}{(4+2B)_H(6-B)_H}}\nonumber\\
&&={\frac{(0)_H(2)_H}{(B)_H(2-B)_H}}
{\frac{(4+B)_H(6+B)_H}{(4+2B)_H(6)_H}}
{\frac{(2)_H(4)_H}{(2+B)_H(4-B)_H}}\nonumber\\
&&~\times{\frac{(2+B)_H(4+B)_H}{(2+2B)_H(4)_H}}
{\frac{(4)_H(6)_H}{(4+B)_H(6-B)_H}}{\frac{(B)_H(2+B)_H}
{(2B)_H(2)_H}}\nonumber\\
&&=\{1\}_0\{5+B\}_0\{3\}_0\{3+B\}_0\{5\}_0\{1+B\}_0\nonumber\\
&&=\{1\}_0\{H-1\}_0\{3\}_0\{H-3\}_0\{5\}_0\{H-5\}_0=[1]_0[3]_0[5]_0.
\Enar
Note that to show above we made replacements at the very basic building
block level. We have added some extra terms in the numerators and 
denominators which eventually cancel. For  making replacements, 
although we do not have some universal rules but certain patterns
can be observed within a theory. In the above example, the second block
of each unit is replaced by a linear function of $h$ with 
integer coefficients, for example $(2)$ by $(h-4)$ and $(3)$ by $(h-3)$
and so on.

One can verify that expression \rref{datpram} holds for any 
$a_{2n-1}^{(1)}\rightarrow c_n^{(1)}$,
so that one can write
\Bear
&S_{ab}^{c_n^{(1)}}&=S_{ab}^{a_{2n-1}^{(1)}}
S_{a{\bar b}}^{a_{2n-1}^{(1)}}|_{h\rightarrow H}
\nonumber\\
&&=\prod_{{p=|a-b|+1}\atop{\rm step 2}}^{{}\atop{a+b-1}}
\{p\}\prod_{{p=|a-b|+1}\atop{\rm step 2}}^{{}\atop{a+b-1}}
\{h-p\}|_{h\rightarrow H}\nonumber\\
&&=\prod_{{p=|a-b|+1}\atop{\rm step 2}}^{{}\atop{a+b-1}}\{p\}_0
\prod_{{p=|a-b|+1}\atop{\rm step 2}}^{{}\atop{a+b-1}}\{H-p\}_0\nonumber\\
&&=\prod_{{p=|a-b|+1}\atop{\rm step 2}}^{{}\atop{a+b-1}}[p]_0,
\Enar
and which is the same as the one given in expression (5.3) of Ref. \cite{CDS}.

Alternatively one can also write $S$-matrices according to 
\rref{nfora}--\rref{nfor}, 
after cancelling zeros and poles, in the following fashion.
\Bear
&S_{ab}^{c_n^{(1)}}=S_{ab}^{a_{2n-1}^{(1)}}
S_{a{\bar b}}^{a_{2n-1}^{(1)}}|_{h\rightarrow H},
\qquad a,b=1,2,....,n-1,\nonumber\\
&S_{an}^{c_n^{(1)}}=S_{an}^{a_{2n-1}^{(1)}}|_{h\rightarrow H},\nonumber\\
&S_{nn}^{c_n^{(1)}}=S_{nn}^{a_{2n-1}^{(1)}}|_{h\rightarrow H}.
\Enar
In this case one would arrive at the expressions given in 
(5.4) of Ref. \cite{CDS}.


\section{$e_6^{(1)}\rightarrow f_4^{(1)}$}
\setcounter{equation}{0}

The folding of $e_6^{(1)}$ Dynkin diagram with labelling is 
already given in \rref{ef}.
In this case the Coxeter number of $e_6^{(1)}$ is $12$, whereas the 
renormalised Coxeter number for the dual pair 
($f_4^{(1)}$, $e_6^{(2)}$) according to the 
notations of Ref. \cite{CDS} is
$H=12+3B$. $S$-matrices of $e_6^{(1)}$ can be found in Table 1 of 
Ref. \cite{BCDSc}.
We have,
\beq
S_{ab}^{f_4^{(1)}}=S_{ab}^{e_6^{(1)}}
S_{a{\bar b}}^{e_6^{(1)}}|_{h\rightarrow H},
\eeq              
and $S_{11}^{f_4^{(1)}}$ and $S_{21}^{f_4^{(1)}}$ are given as
\Bear
&S_{11}^{f_4^{(1)}}&=S_{11}^{e_6^{(1)}}
S_{1{\bar 1}}^{e_6^{(1)}}|_{h\rightarrow H}
=\{1\}\{7\}\{5\}\{11\}|_{h\rightarrow H}=\{1\}\{2h/3-1\}\{h/3+1\}\{h-1\}
|_{h\rightarrow H}\nonumber\\
&&=\{1\}_0\{2H/3-1\}_0\{H/3+1\}_0\{H-1\}_0=[1]_0[H/3+1]_0,\nonumber\\
&S_{21}^{f_4^{(1)}}&=S_{21}^{e_6^{(1)}}
S_{2{\bar 1}}^{e_6^{(1)}}|_{h\rightarrow H}
=4~4|_{h\rightarrow H}=\{4\}\{8\}\{4\}\{8\}|_{h\rightarrow H}\nonumber\\   
&&=\{h/6+2\}\{5h/6-2\}\{h/2-2\}\{h/2+2\}|_{h\rightarrow H}\nonumber\\
&&=\{H/6+2\}_0\{5H/6-2\}_0\{H/2-2\}_0\{H/2+2\}_0=[H/6+2]_0[H/2+2]_0.
\Enar 
The above $S$-matrices appear after the expression (4.5) in 
Ref. \cite{CDS}.
The rest of the $S$-matrices are calculated in appendix A (A.1-A.8).
Again one can show that \rref{nfora}--\rref{nfor} holds for 
$S$-matrices. For obtaining
those one has to write down $S$-matrices (after cancelling zeros and poles
present in the above ones, see expressions (4.6) of Ref. \cite{CDS}) 
in terms of 
elementary building blocks, $(x)$, and replace $x$ in terms of linear
function of $h$ appropriately 
there itself. To demonstrate this we take a sample $S$-matrix 
element ($S_{12}$) here,
rest of the calculations can be found in the appendix A (A.9-A.14).

\Bear
&S_{12}^{f_4^{(1)}}&=S_{12}^{e_6^{(1)}}|_{h\rightarrow H}
=4|_{h\rightarrow H}=\{4\}\{8\}|_{h\rightarrow H}
={\frac{(3)(5)}{(3+B)(5-B)}}{\frac{(7)(9)}{(7+B)(9-B)}}
{\Big |}_{h\rightarrow H}\nonumber\\
&&={\frac{(h/6+1)(h/2-1)}{(h/2-3+B)(h/6+3-B)}}{\frac{(h/2+1)(5h/6-1)}
{(5h/6-3+B)(h/2+3-B)}}{\Big |}_{h\rightarrow H}\nonumber\\
&&=\hcd{h/6+1}\hcd{h/2+1}|_{h\rightarrow H}=
\hcd{H/6+1}_0\hcd{H/2+1}_0\nonumber\\
&&={\frac{(3+B/2)_H(5+3B/2)_H}{(3+5B/2)_H(5-B/2)_H}}
{\frac{(7+3B/2)_H(9+5B/2)_H}
{(7+7B/2)_H(9+B/2)_H}}\nonumber\\
&&=\{4+B\}_{1/2}\{8+2B\}_{1/2}
=\{H/3\}_{1/2}\{2H/3\}_{1/2}=[H/3]_{1/2}.
\Enar 
Where we have used  modified blocks $\hcd{x}$ and $\hcd{x}_0$, given by
\beq
\hcd{x}={\frac{(x)(h/3-2+x)}{(h/3-4+x+B)(x+2-B)}},
\qquad \hcd{x}_0={\frac{(x)_H(H/3-2+x)_H}{(H/3-4+x+B)_H(x+2-B)_H}}.
\eeq

\section{$d_{n+1}^{(1)}\rightarrow b_n^{(1)}$}
\setcounter{equation}{0}

Folding of Dynkin diagram and mass labellings are given in the following 
figure.

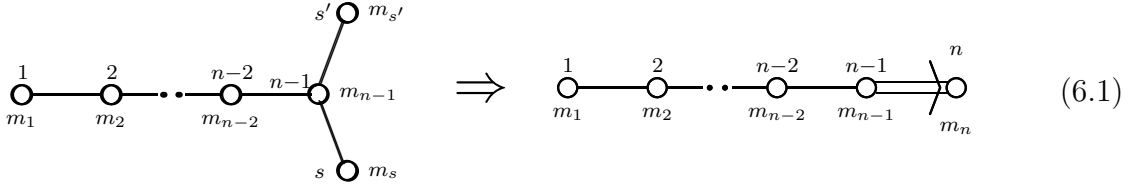
\beq
\dddnut{1}.{2}.{{n-2}}.{{n-1}}.{s}.{{s'}}.
\llap{\dddndt{m_1}.{m_2}.{m_{n-2}}.{m_{n-1}}.{m_s}.{m_{s'}}.}
\Rightarrow
\ddbnidr{m_1}.{m_2}.{m_{n-2}}.{m_{n-1}}.{m_n}.
\llap{\ddbniur{1}.{2}.{{n-2}}.{{n-1}}.{n}.}
\label{dtb}
\eeq

Again to make life simple we take a specific example of this 
family and show the results. It can be seen that  generalisation 
will follow immediately.
Here we discuss the case $d_{5}^{(1)}\rightarrow b_4^{(1)}$. Apart from
the particles 1, 2 and 3 in $d_{5}^{(1)}$ one has mass degenerate
conjugate pair of particles $s$ and $\bar s$. After the folding 
degeneracy is lost and $4^{\rm th}$ particle of $b_4^{(1)}$ is a linear
combination of $s$ and $\bar s$.\footnote{Notice that in $d_{n+1}^{(1)}$,
$n={\rm odd}$, one has mass degenerate pair $s$ and $s'$ which are self 
conjugate. But the folding $d_{n+1}^{(1)} \rightarrow b_n^{(1)}$, 
identifies the particles $s$ and $s'$ to produce $n^{\rm th}$ 
particle of $b_n^{(1)}$.
In this case we write $S_{nn}^{b_n^{(1)}}$ as,
\beq
S_{nn}^{b_n^{(1)}}=S_{ss}^{d_{n+1}^{(1)}}S_{ss'}^{d_{n+1}^{(1)}},
\eeq
which is analogous to \rref{nfora}. In the next section a similar 
situation arises again when three
self conjugate particles of $d_4^{(1)}$ are identified to produce 
$g_2^{(1)}$ theory. There too one should understand the 
formula \rref{nfora} accordingly.} In this case first we 
show the formulae \rref{nfora}--\rref{nfor} and postpone the 
discussion of \rref{datpram} till the end of this section. 
The Coxeter number for $d_{5}^{(1)}$ is 8 and
the renormalised Coxeter for the pair ($b_4^{(1)}$, $a_{7}^{(2)}$) 
is $8-B/2$.
According to the expression \rref{nfora}--\rref{nfor},
$a,~b =1,2,3$ and $m,~n=4$.  So,

\Bear
&S_{44}^{b_4^{(1)}}&=S_{ss}^{d_5^{(1)}}S_{s\bar s}^{d_5^{(1)}}
|_{h\rightarrow H}
=\{1\}\{5\}\{3\}\{7\}\nonumber\\
&&={\frac{(0)(2)}{(B)(2-B)}}
{\frac{(4)(6)}{(4+B)(6-B)}}{\frac{(2)(4)}{(2+B)(4-B)}}
{\frac{(6)(8)}{(6+B)(8-B)}}{\Big |}_{h\rightarrow H}\nonumber\\
&&={\frac{(0)(h-6)}{(h-8+B)(2-B)}}
{\frac{(4)(h-2)}{(h-4+B)(6-B)}}\nonumber\\
&&~\times{\frac{(2)(h-4)}{(h-6+B)(4-B)}}
{\frac{(6)(h)}{(h-2+B)(8-B)}}{\Big |}_{h\rightarrow H}\nonumber\\
&&={\frac{(0)_H(H-6)_H}{(H-8+B)_H(2-B)_H}}
{\frac{(4)_H(H-2)_H}{(H-4+B)_H(6-B)_H}}\nonumber\\
&&~\times{\frac{(2)_H(H-4)_H}{(H-6+B)_H(4-B)_H}}
{\frac{(6)_H(H)_H}{(H-2+B)_H(8-B)_H}}\nonumber\\
&&={\frac{(0)_H(2-B/2)_H}{(B/2)_H(2-B)_H}}
{\frac{(4)_H(6-B/2)_H}{(4+B/2)_H(6-B)_H}}\nonumber\\
&&~\times{\frac{(2)_H(4-B/2)_H}
{(2+B/2)_H(4-B)_H}}
{\frac{(6)_H(8-B/2)_H}{(6+B/2)_H(8-B)_H}}\nonumber\\
&&=\{1-B/4\}_{-1/4}\{5-B/4\}_{-1/4}\{3-B/4\}_{-1/4}
\{7-B/4\}_{-1/4}\nonumber\\
&&=\{H/2-3\}_{-1/4}\{H/2+1\}_{-1/4}\{H/2-1\}_{-1/4}
\{H/2+3\}_{-1/4}\nonumber\\
&&=\prod_{{p=-3}\atop{\rm step 2}}^{{}\atop{3}}\{H/2-p\}_{1/4}.
\label{for4}
\Enar
For $S_{ab}^{b_4^{(1)}}~{\rm and}~ S_{a4}^{b_4^{(1)}},~a,~b=1,~2,~3$ 
matrix elements we begin with the matrix
elements of $d_{5}^{(1)}$ as given in the expressions (4.13) and (4.16) of 
Ref. \cite{BCDSc}, respectively.
\Bear
&S_{ab}^{b_4^{(1)}}&=S_{ab}^{d_5^{(1)}}|_{h\rightarrow H}\nonumber\\
&&=\prod_{{p=|a-b|+1}\atop{\rm step 2}}^{{}\atop{a+b-1}}\{p\}\{h-p\}
|_{h\rightarrow H}\nonumber\\
&&=\prod_{{p=|a-b|+1}\atop{\rm step 2}}^{{}\atop{a+b-1}}
\{p\}_0\{H-p\}_0\nonumber\\
&&=\prod_{{p=|a-b|+1}\atop{\rm step 2}}^{{}\atop{a+b-1}}[p]_0,
\Enar
and
\Bear
&S_{a4}^{b_4^{(1)}}&=S_{as}^{d_5^{(1)}}|_{h\rightarrow H}\nonumber\\
&&=\prod_{{p=0}\atop{\rm step 2}}^{{}\atop{2a-2}}\{5-a+p\}
|_{h\rightarrow H}\nonumber\\
&&=\prod_{{p=0}\atop{\rm step 2}}^{{}\atop{2a-2}}\{h/2+1-a+p\}
|_{h\rightarrow H}
=\prod_{{p=1}\atop{\rm step 2}}^{{}\atop{2a-1}}\{h/2-a+p\}
|_{h\rightarrow H}\nonumber\\
&&=\prod_{{p=1}\atop{\rm step 2}}^{{}\atop{2a-1}}\{H/2-a+p\}_0.
\Enar

To show the formula \rref{datpram}, one has to go
to the elementary building block level and make substitutions their
itself. Of course $S_{44}$ is the same as expression \rref{for4},
for the rest one proceeds in the following manner,
\Bear
&S_{14}^{b_4^{(1)}}&=S_{1s}^{d_5^{(1)}}S_{1\bar s}^{d_5^{(1)}}
|_{h\rightarrow H}\nonumber\\
&&=\{4\}\{4\}|_{h\rightarrow H}={\frac{(3)(5)}{(3+B)(5-B)}}
{\frac{(3)(5)}{(3+B)(5-B)}}|_{h\rightarrow H}\nonumber\\
&&={\frac{(h/2-1)(3h/2-7)}{(3h/2-9+B)(h/2+1-B)}}
{\frac{(7-h/2)(h/2+1)}{(h/2-1+B)(9-h/2-B)}}|_{h\rightarrow H}\nonumber\\
&&={\frac{(H/2-1)_H(3H/2-7)_H}{(3H/2-9+B)_H(H/2+1-B)_H}}
{\frac{(7-H/2)_H(H/2+1)_H}{(H/2-1+B)_H(9-H/2-B)_H}}\nonumber\\
&&={\frac{(H/2-1)_H(5-3B/4)_H}{(3+B/4)_H(H/2+1-B)_H}}
{\frac{(3+B/4)_H(H/2+1)_H}{(H/2-1+B)_H(5-3B/4)_H}}\nonumber\\
&&={\frac{(H/2-1)_H(H/2+1)_H}{(H/2-1+B)_H(H/2+1-B)_H}}=\{H/2\}_0.
\Enar
Remaining ones are evaluated in a similar fashion. Caculations are 
straightforward but tedious so we do not produce them here.


\section{$d_4^{(1)}\rightarrow g_2^{(1)}$}
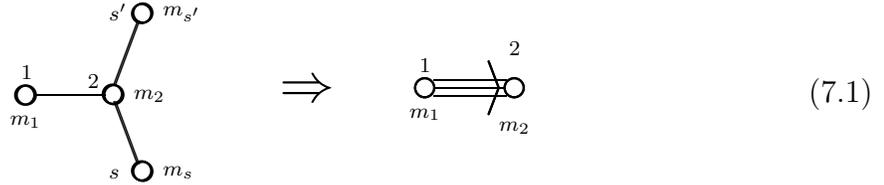
\setcounter{equation}{0}

\beq
\dduf{1}.{2}.{s}.{s'}.
\llap{\dddf{m_1}.{m_2}.{m_s}.{m_{s'}}.}
\quad\Rightarrow\quad \ddgiid{m_1}.{m_2}.
\llap{\ddgiiu{1}.{2}.}
\eeq

The case of $g_2^{(1)}$ is a little different from the rest.
There are three mass degenerate self conjugate particles, viz.
$m_1,~m_s,~m_{s'}$ in $d_4^{(1)}$.
In this case 
three points of the parent theory are identified in the process of
folding unlike previous cases where at most two points were identified.
For the dual pair ($g_2^{(1)}$, $d_4^{(3)}$) the renormalised Coxeter
$H=6+3B$ and the Coxeter for $d_4^{(1)}$ is 6.
In this we show the \rref{nfora}--\rref{nfor} for the $S$-matrix elements.
The expression \rref{datpram} can be shown without much difficulty and 
will not be produced here. 
To show \rref{nfora}--\rref{nfor} we proceed followingly.
\Bear
&S_{11}^{g_2}&=S_{11}^{d_4}S_{1s}^{d_4}|_{h\rightarrow H}
=S_{ss}^{d_4}S_{ss'}^{d_4}|_{h\rightarrow H}
=S_{s's'}^{d_4}S_{s'1}^{d_4}|_{h\rightarrow H}
=\{1\}\{5\}\{3\}|_{h\rightarrow H}\nonumber\\
&&={\frac{(0)(2)}{(B)(2-B)}}{\frac{(4)(6)}{(4+B)(6-B)}}
{\frac{(2)(4)}{(2+B)(4-B)}}
{\Big |}_{h\rightarrow H}\nonumber\\
&&={\frac{(0)(2)}{(B)(2-B)}}{\frac{(h-2)(h)}
{(h-2+B)(h-B)}}{\frac{(h/3)(2h/3)}
{(2h/3-2+B)(h/3+2-B)}}{\Big |}_{h\rightarrow H}\nonumber\\
&&={\frac{(0)_H(2)_H}{(B)_H(2-B)_H}}{\frac{(H-2)_H(H)_H}
{(H-2+B)_H(H-B)_H}}{\frac{(H/3)_H(2H/3)_H}
{(2H/3-2+B)_H(H/3+2-B)_H}}\nonumber\\
&&=\{1\}_0\{H-1\}_0\{H/2\}_{1/2},\nonumber\\
&S_{12}^{g_2}&=S_{12}^{d_4}|_{h\rightarrow H}
=\{2\}\{4\}|_{h\rightarrow H}
={\frac{(1)(3)}{(1+B)(3-B)}}{\frac{(3)(5)}{(3+B)(5-B)}}
{\Big |}_{h\rightarrow H}\nonumber\\
&&={\frac{(1)(2h/3-1)}{(2h/3-3+B)(3-B)}}{\frac{(h/3+1)(h-1)}
{(h-3+B)(h/3+3-B)}}{\Big |}_{h\rightarrow H}\nonumber\\
&&={\frac{(1)_H(2H/3-1)_H}{(2H/3-3+B)_H(3-B)_H}}{\frac{(H/3+1)_H(H-1)_H}
{(H-3+B)_H(H/3+3-B)_H}}\nonumber\\
&&={\frac{(1)_H(2H/3-1)_H}{(H-5)_H(5-H/3)_H}}{\frac{(H/3+1)_H(H-1)_H}
{(4H/3-5)_H(5)_H}}\nonumber\\
&&=\{H/3\}_1\{2H/3\}_1,\nonumber\\
&S_{22}^{g_2}&=S_{22}^{d_4}|_{h\rightarrow H}
=1~3|_{h\rightarrow H}=\{1\}\{5\}\{3\}\{3\}|_{h\rightarrow H}\nonumber\\
&&={\frac{(0)(2)}{(B)(2-B)}}{\frac{(4)(6)}{(4+B)(6-B)}}\nonumber\\
&&~\times{\frac{(2)(4)}{(2+B)(4-B)}}{\frac{(2)(4)}{(2+B)(4-B)}}
{\Big |}_{h\rightarrow H}\nonumber\\
&&={\frac{(0)(2h/3-2)}{(2h/3-4+B)(2-B)}}
{\frac{(h/3+2)(h)}{(h-2+B)(h/3+4-B)}}\nonumber\\
&&~\times{\frac{(2)(2h/3)}{(2h/3-2+B)(4-B)}}
{\frac{(h/3)(h-2)}{(h-4+B)(h/3+2-B)}}{\Big |}_{h\rightarrow H}\nonumber\\
&&={\frac{(0)_H(2H/3-2)_H}{(2H/3-4+B)_H(2-B)_H}}
{\frac{(H/3+2)_H(H)_H}{(H-2+B)_H(H/3+4-B)_H}}\nonumber\\
&&~\times{\frac{(2)_H(2H/3)_H}{(2H/3-2+B)_H(4-B)_H}}
{\frac{(H/3)_H(H-2)_H}{(H-4+B)_H(H/3+2-B)_H}}\nonumber\\
&&={\frac{(0)_H(2H/3-2)_H}{(H-6)_H(4-H/3)_H}}{\frac{(H/3+2)_H(H)_H}
{(4H/3-6)_H(6)_H}}\nonumber\\
&&~\times{\frac{(2)_H(2H/3)_H}{(H-4)_H(6-H/3)_H}}
{\frac{(H/3)_H(H-2)_H}{(4H/3-6)_H(4)_H}}
\nonumber\\
&&=\{H/3-1\}_1\{2H/3+1\}_1\{H/3+1\}_1\{2H/3-1\}_1.
\Enar
\section{Twisted theories}
\setcounter{equation}{0}
For the dual theories, $d_{n+1}^{(2)},~e_6^{(2)},~d_4^{(3)}$ and
$a_{2n-1}^{(2)}$
the story is a little different. These are twisted theories obtained 
by exploiting the symmetries of extended Dynkin diagrams of simply 
laced theories. In these cases as described in the Ref. \cite{BCDSc} 
the mass spectrum of the
daughter theory is just a subset of that of parent theory.
In the same way (as mentioned at the end of the section 3) 
we notice that the $S$-matrices of these theories are
just subsets of the corresponding parent theories with the Coxeter $h$
replaced by the renormalised $H$.

\subsection{$d_{n+2}^{(1)}\rightarrow d_{n+1}^{(2)}$}
Here two degenerate particles $s~{\rm and}~s'(s~{\rm and}~{\bar s}~{\rm
for}~d_{\rm odd}^{(1)} )$ are lost.
All other particles survive folding and labelled by $1,2,..,n$.
The Coxeter number $h$ for $d_{n+2}^{(1)}$ theory is $2n+2$, whereas the
renormalized Coxeter for the pair ($d_{n+1}^{(2)}$, $c_n^{(1)}$),
$H=2n+B$.
$S$-matrices are calculated as (for $d_n^{(1)}$ $S$-matrices, see
expression (4.13) of Ref. \cite{BCDSc} ),
\Bear
&S_{ab}^{d_{n+1}^{(2)}}&=S_{ab}^{d_{n+2}^{(1)}},\qquad a,b=1,2,..,n.
\nonumber\\
&&=\prod_{{p=|a-b|+1}\atop{\rm step 2}}^{{}\atop{a+b-1}}
\{p\}\{h-p\}|_{h\rightarrow H}\nonumber\\
&&=\prod_{{p=|a-b|+1}\atop{\rm step 2}}^{{}\atop{a+b-1}}\{p\}_0
\{H-p\}_0\nonumber\\
&&=\prod_{{p=|a-b|+1}\atop{\rm step 2}}^{{}\atop{a+b-1}}[p]_0.
\Enar
These are the same $S$-matrices obtained in the expression (4.5)
of section 4 for the dual theory $c_n^{(1)}$.

\subsection{$e_7^{(1)}\rightarrow e_6^{(2)}$} 

In this case masses $m_2,~m_4,~m_5,~{\rm and}~m_7$ of $e_7^{(1)}$
survive and become particle 1,2,3 and 4 respectively 
in $e_6^{(2)}$ after relabelling.
Here $h=18$ for $e_7^{(1)}$ and $H=12+3B$ for the dual pair
($e_6^{(2)}$, $f_4^{(1)}$).
$S$-matrices for $e_6^{(2)}$ are given in Ref. \cite{CDS} 
(following equation (4.5), before 
cancelling zeros and poles) can be obtained very easily from the
$e_7^{(1)}$ $S$-matrices shown in Table 2 of the Ref. \cite{BCDSc}
as follows.
Here we give just the $S_{11}^{e_6^{(2)}}$ rest of the calculations are
presented in appendix B.

\Bear
&S_{11}^{e_6^{(2)}}&=S_{22}^{e_7^{(1)}}|_{h\rightarrow H}
=1~7|_{h\rightarrow H}=\{1\}\{17\}\{7\}\{11\}|_{h\rightarrow H}\nonumber\\
&&=\{1\}\{h-1\}\{h/3+1\}\{2h/3-1\}|_{h\rightarrow H}\nonumber\\
&&=\{1\}_0\{H-1\}_0\{H/3+1\}_0\{2H/3-1\}_0\nonumber\\
&&=[1]_0[H/3+1]_0.\nonumber\\
\Enar
Again notice that these $S$-matrices match with the $S$-matrices
of the dual theory, $f_4^{(1)}$, obtained in section 5 (expressions
(5.2) and (A.1--A.8)).

\subsection{$e_6^{(1)}\rightarrow d_4^{(3)}$} 
 In this case the particles $2$ and $4$ of $e_6^{(1)}$ survive and 
relabelled as particle 1 and 2 in $d_4^{(3)}$. 
The Coxeter number $h=12$ for $e_6^{(1)}$ and the renormalized Coxeter,
as mentioned earlier, for the pair 
($d_4^{(3)}$,  $g_2^{(1)}$) is $6+3B$. The corresponding $S$-
matrix elements,

\Bear
&S_{11}^{d_4^{(3)}}&=S_{22}^{e_6^{(1)}}|_{h\rightarrow H}
=1~5|_{h\rightarrow H}=\{1\}\{11\}\{5\}\{7\}|_{h\rightarrow H}\nonumber\\
&&=\{1\}\{h-1\}\{h/3+1\}\{2h/3-1\}|_{h\rightarrow H}\nonumber\\
&&=\{1\}_0\{H-1\}_0\{H/3+1\}_0\{2H/3-1\}_0\nonumber\\
&&=\{1\}_0\{5+3B\}_0\{3+B\}_0\{3+2B\}_0\nonumber\\
&&={\frac{(0)_H(2)_H}{(B)_H(2-B)_H}}
{\frac{(4+3B)_H(6+3B)_H}{(4+4B)_H(6+2B)_H}}\nonumber\\
&&~\times{\frac{(2+B)_H(4+B)_H}{(2+2B)_H(4)_H}}
{\frac{(2+2B)_H(4+2B)_H}{(2+3B)_H(4+B)_H}}\nonumber\\
&&={\frac{(0)_H(2)_H}{(B)_H(2-B)_H}}{\frac{(4+3B)_H(6+3B)_H}
{(4+4B)_H(6+2B)_H}}
{\frac{(2+B)_H(4+2B)_H}{(2+3B)_H(4)_H}}\nonumber\\
&&={\frac{(0)_H(2)_H}{(H/3-2)_H(4-H/3)_H}}{\frac{(H-2)_H(H)_H}
{(2+2H/3)_H(4H/3-4)_H}}{\frac{(H/3)_H(2H/3)_H}{(H-4)_H(4)_H}}\nonumber\\
&&=\{1\}_0\{H-1\}_0\{H/2\}_{1/2}.\nonumber\\
\Enar
Similarly,
\Bear
&S_{12}^{d_4^{(3)}}&=S_{24}^{e_6^{(1)}}|_{h\rightarrow H}
=2~4~6|_{h\rightarrow H}=\{2\}\{10\}\{4\}\{8\}\{6\}\{6\}
|_{h\rightarrow H}\nonumber\\
&&=\{2\}\{h-2\}\{h/3\}\{2h/3\}\{h/3+2\}\{2h/3-2\}
|_{h\rightarrow H}\nonumber\\
&&=\{2\}_0\{H-2\}_0\{H/3\}_0\{2H/3\}_0\{H/3+2\}_0\{2H/3-2\}_0\nonumber\\
&&=[2]_0[H/3]_0[H/3+2]_0=\{H/3\}_1\{2H/3\}_1,\nonumber\\
&S_{22}^{d_4^{(3)}}&=S_{44}^{e_6^{(1)}}|_{h\rightarrow H}
=1~3^2~5^3|_{h\rightarrow H}=\{1\}\{h-1\}\{3\}\{h-3\}\{h/3-1\}
\{2h/3+1\}\nonumber\\
&&~\times\{2h/3-3\}\{h/3+3\}\{h/3+1\}^2\{2h/3-1\}^2
|_{h\rightarrow H}\nonumber\\
&&=\{1\}_0\{H-1\}_0\{3\}_0\{H-3\}_0\{H/3-1\}_0\{2H/3+1\}_0\nonumber\\
&&~\times\{2H/3-3\}_0\{H/3+3\}_0\{H/3+1\}_0^2\{2H/3-1\}_0^2\nonumber\\
&&=[1]_0[3]_0[H/3-1]_0[H/3+3]_0[H/3+1]_0^2\nonumber\\
&&=\{H/3-1\}_1\{2H/3+1\}_1\{H/3+1\}_1\{2H/3-1\}_1,
\Enar
where in the last line of each element we have omitted few steps 
and given simplified result. Results are again same as the ones 
obtained in the section 7 for the untwisted dual pair $g_2^{(1)}$.

\subsection{$d_{2n}^{(1)}\rightarrow a_{2n-1}^{(2)}$}
Here we consider the example of $d_6^{(1)}\rightarrow a_5^{(2)}$, and 
show the result for just one element viz. $S_{33}^{a_5^{(2)}}$. For the 
rest we don't have the results yet, but we have strong hope that 
they can be manipulated in the similar fashion. For $d_6^{(1)}$ 
the Coxeter $h=10$ and the renormalised Coxeter for 
the pair ($a_5^{(2)}$, $b_3^{(1)}$) is $6-B/2$.
\Bear
&S_{33}^{a_5^{(2)}}&=S_{ss}^{d_6^{(1)}}|_{h\rightarrow H}
=\{1\}\{9\}\{5\}|_{h\rightarrow H}\nonumber\\
&&={\frac{(0)(2)}{(B)(2-B)}}{\frac{(8)(10)}{(8+B)(10-B)}}
{\frac{(4)(6)}{(4+B)(6-B)}}{\Big |}_{h\rightarrow H}\nonumber\\
&&={\frac{(0)(h-8)}{(h-10+B)(2-B)}}{\frac{(8)(h)}{(h-2+B)(10-B)}}
{\frac{(4)(h-4)}{(h-6+B)(6-B)}}{\Big |}_{h\rightarrow H}\nonumber\\
&&={\frac{(0)(H-8)}{(H-10+B)(2-B)}}{\frac{(8)(H)}{(H-2+B)(10-B)}}
{\frac{(4)(H-4)}{(H-6+B)(6-B)}}\nonumber\\
&&={\frac{(0)(-2-B/2)}{(-4+B/2)(2-B)}}{\frac{(-10+B)(6-B/2)}{(4+B/2)(-8)}}
{\frac{(4)(2-B/2)}{(B/2)(6-B)}}\nonumber\\
&&={\frac{(0)(4-B/2)}{(2+B/2)(2-B)}}{\frac{(2)(6-B/2)}{(4+B/2)(4-B)}}
{\frac{(4)(2-B/2)}{(B/2)(6-B)}}\nonumber\\
&&={\frac{(0)(2-B/2)}{(B/2)(2-B)}}{\frac{(2)(4-B/2)}{(2+B/2)(4-B)}}
{\frac{(4)(6-B/2)}{(4+B/2)(6-B)}}\nonumber\\
&&=\{1-B/4\}_{-1/4}\{3-B/4\}_{-1/4}\{5-B/4\}_{-1/4}\nonumber\\
&&=\{H/2-2\}_{-1/4}\{H/2\}_{-1/4}\{H/2+2\}_{-1/4}\nonumber\\
&&=\prod_{{p=-2}\atop{{\rm step} 2}}^{{}\atop{2}}\{H/2-p\}_{-1/4}.
\Enar
\section{Summary and discussion}
\setcounter{equation}{0}
In the present paper we made an attempt to extend the idea of folding and
classical reduction of Toda field theories to the quantum case. We
have shown that with the help of simple formulae viz. 
\rref{datpram}--\rref{nfor} one can construct the exact quantum 
$S$-matrices of the non-simply laced Toda field theories from the
$S$-matrices of simply laced Toda theories by just replacing
the Coxeter number appropriately. Here we concentrated on the
$S$-matrices, but fact is that this idea of folding also works
for the three point couplings.\footnote{For mass ratios folding 
will work in a straight forward manner and one will have floating 
mass ratios like the 
ones mentioned in Refs. \cite{CDS, DGZc}.} Couplings of non-simply
laced theories can be evaluated from the couplings of simply laced
theories by replacing the Coxeter number accordingly.
For example, let us consider the coupling,
$U_{12}^1=5H/6-1$(in units of $\pi/H$) for the dual pair $f_4^{(1)}$
and $e_6^{(2)}$ (see expression (4.4) in the Ref. \cite{CDS}). We
show in the following that this can be obtained in two ways 
from the parent theories, $e_6^{(1)}$ and $e_7^{(1)}$.
\Bear
{U_{12}^1}^{(f_4^{(1)},~e_6^{(2)})}={C_{12}^1}^{e_6^{(1)}}
|_{h\rightarrow H}
=9|_{h\rightarrow H}=5h/6-1|_{h\rightarrow H}=5H/6-1.
\Enar
Alternatively,
\Bear
{U_{12}^1}^{(f_4^{(1)},~e_6^{(2)})}={C_{24}^2}^{e_7^{(1)}}
|_{h\rightarrow H}
=14|_{h\rightarrow H}=5h/6-1|_{h\rightarrow H}=5H/6-1.
\Enar
This may be one of reasons why folding works for the quantum $S$-matrices.
One point lacking in our formalism is that there is no universal way of 
writing the block $\{x\}$ in terms of $h$. $x$ should be written as
a linear fuction of $h$. Although certain patterns within a theory may
be visible but for clarity one would seek some universal rule. At present
we are investigating along these points. To conclude we think this is a 
new approach and in future will be able to shed some light on structure
of $S$-matrices of Toda field theories.
\section*{Acknowledgements} 
 I would like to thank Prof. Ryu Sasaki for reading  the manuscript 
carefully and for making valuable comments and
suggestions at various stages of the work.

\section*{Appendix A: $e_6^{(1)}\rightarrow f_4^{(1)}$}
\setcounter{equation}{0}
\renewcommand{\theequation}{A.\arabic{equation}}
\Bear
&S_{31}^{f_4^{(1)}}&=S_{31}^{e_6^{(1)}}S_{3{\bar 1}}^{e_6^{(1)}}
|_{h\rightarrow H}
=\{4\}\{6\}\{10\}\{2\}\{6\}\{8\}|_{h\rightarrow H}\nonumber\\
&&=\{h/3\}\{h/3+2\}\{h-2\}\{2\}\{2h/3-2\}\{2h/3\}
|_{h\rightarrow H}\nonumber\\
&&=\{H/3\}_0\{H/3+2\}_0\{H-2\}_0\{2\}_0\{2H/3-2\}_0\{2H/3\}_0\nonumber\\
&&=[2]_0[H/3]_0[H/3+2]_0\\
&S_{41}^{f_4^{(1)}}&=S_{41}^{e_6^{(1)}}
S_{4{\bar 1}}^{e_6^{(1)}}|_{h\rightarrow H}
=3~5~3~5|_{h\rightarrow H}=\{3\}\{9\}\{5\}\{7\}\{3\}\{9\}\{5\}\{7\}
|_{h\rightarrow H}\nonumber\\   
&&=\{3\}\{h-3\}\{h/3+1\}\{2h/3-1\}\{h/3-1\}\nonumber\\
&&~\times\{2h/3+1\}\{2h/3-3\}\{h/3+3\}
|_{h\rightarrow H}\nonumber\\
&&=\{3\}_0\{H-3\}_0\{H/3+1\}_0\{2H/3-1\}_0\nonumber\\
&&~\times\{H/3-1\}_0\{2H/3+1\}_0\{2H/3-3\}_0\{H/3+3\}_0\nonumber\\
&&=[3]_0[H/3+1]_0[H/3-1]_0[H/3+3]_0\\
&S_{22}^{f_4^{(1)}}&=S_{22}^{e_6^{(1)}}S_{22}^{e_6^{(1)}}|_{h\rightarrow H}
=1~5~1~5|_{h\rightarrow H}=\{1\}\{11\}\{5\}\{7\}\{1\}\{11\}\{5\}\{7\}
|_{h\rightarrow H}\nonumber\\   
&&=\{1\}\{h-1\}\{h/3+1\}\{2h/3-1\}\{h/3-3\}\nonumber\\
&&~\times\{2h/3+3\}\{2h/3-3\}\{h/3+3\}
|_{h\rightarrow H}\nonumber\\
&&=\{1\}_0\{H-1\}_0\{H/3+1\}_0\{2H/3-1\}_0\{H/3-3\}_0\nonumber\\
&&~\times\{2H/3+3\}_0\{2H/3-3\}_0\{H/3+3\}_0\nonumber\\
&&=[1]_0[H/3+1]_0[H/3-3]_0[H/3+3]_0\\
&S_{23}^{f_4^{(1)}}&=S_{23}^{e_6^{(1)}}
S_{2\bar 3}^{e_6^{(1)}}|_{h\rightarrow H}
=3~5~3~5|_{h\rightarrow H}=\{3\}\{9\}\{5\}\{7\}\{3\}\{9\}\{5\}\{7\}
|_{h\rightarrow H}\nonumber\\   
&&=\{h/6+1\}\{5h/6-1\}\{h/6+3\}\{5h/6-3\}\nonumber\\
&&~\times\{h/2-3\}\{h/2+3\}\{h/2-1\}\{h/2+1\}
|_{h\rightarrow H}\nonumber\\
&&=\{H/6+1\}_0\{5H/6-1\}_0\{H/6+3\}_0\{5H/6-3\}_0\nonumber\\
&&~\times\{H/2-3\}_0\{H/2+3\}_0\{H/2-1\}_0\{H/2+1\}_0\nonumber\\
&&=[H/6+1]_0[H/6+3]_0[H/2+3]_0[H/2+1]_0\\
&S_{24}^{f_4^{(1)}}&=S_{24}^{e_6^{(1)}}S_{24}^{e_6^{(1)}}|_{h\rightarrow H}
=2~4~6~2~4~6|_{h\rightarrow H}\nonumber\\
&&=\{2\}\{10\}\{4\}\{8\}\{6\}\{6\}\{2\}\{10\}\{4\}\{8\}\{6\}\{6\}
|_{h\rightarrow H}\nonumber\\   
&&=\{h/6\}\{5h/6\}\{h/6+2\}\{5h/6-2\}\{h/6+4\}\{5h/6-4\}\nonumber\\
&&~\times\{h/2-4\}\{h/2+4\}\{h/2-2\}\{h/2+2\}\{h/2\}\{h/2\}
|_{h\rightarrow H}\nonumber\\
&&=\{H/6\}_0\{5H/6\}_0\{H/6+2\}_0\{5H/6-2\}_0\{H/6+4\}_0
\{5H/6-4\}_0\nonumber\\
&&~\times\{H/2-4\}_0\{H/2+4\}_0\{H/2-2\}_0\{H/2+2\}_0
\{H/2\}_0\{H/2\}_0\nonumber\\
&&=[H/6]_0[H/6+2]_0[H/6+4]_0[H/2+4]_0[H/2+2]_0[H/2]_0\\
&S_{33}^{f_4^{(1)}}&=S_{33}^{e_6^{(1)}}
S_{3\bar 3}^{e_6^{(1)}}|_{h\rightarrow H}
=\{1\}\{3\}\{5\}\{7\}^2\{9\}\{3\}\{5\}^2\{7\}\{9\}\{11\}
|_{h\rightarrow H}\nonumber\\   
&&=\{1\}\{3\}\{2h/3-3\}\{2h/3-1\}^2\{2h/3+1\}\{h/3-1\}\nonumber\\
&&~\times\{h/3+1\}^2\{h/3+3\}\{h-3\}\{h-1\}|_{h\rightarrow H}\nonumber\\
&&=\{1\}_0\{3\}_0\{2H/3-3\}_0\{2H/3-1\}_0^2
\{2H/3+1\}_0\{H/3-1\}_0\nonumber\\
&&~\times\{H/3+1\}_0^2\{H/3+3\}_0\{H-3\}_0\{H-1\}_0\nonumber\\
&&=[1]_0[3]_0[H/3-1]_0[H/3+1]_0^2[H/3+3]_0\\
&S_{43}^{f_4^{(1)}}&=S_{43}^{e_6^{(1)}}
S_{4\bar 3}^{e_6^{(1)}}|_{h\rightarrow H}
=2~4^2~6~2~4^2~6|_{h\rightarrow H}\nonumber\\
&&=\{2\}\{10\}\{4\}^2\{8\}^2\{6\}\{6\}\{2\}\{10\}\{4\}^2\{8\}^2
\{6\}\{6\}|_{h\rightarrow H}\nonumber\\   
&&=\{2\}\{h-2\}\{h/3\}^2\{2h/3\}^2\{h/3+2\}\{2h/3-2\}\nonumber\\
&&~\times\{h/3-2\}^2\{2h/3+2\}\{4\}\{h-4\}\{2h/3-4\}^2\nonumber\\
&&~\times\{h/3+4\}\{h/3+2\}\{h/3-2\}|_{h\rightarrow H}\nonumber\\
&&=\{2\}_0\{H-2\}_0\{H/3\}_0^2\{2H/3\}_0^2\{H/3+2\}_0
\{2H/3-2\}_0\{H/3-2\}_0^2\nonumber\\
&&~\times\{2H/3+2\}_0\{4\}_0\{H-4\}_0\{2H/3-4\}_0^2\{H/3+4\}_0
\{H/3+2\}_0\{H/3-2\}_0\nonumber\\
&&=[2]_0[4]_0[H/3-2]_0[H/3]_0^2[H/3+2]_0^2[H/3+4]_0\\
&S_{44}^{f_4^{(1)}}&=S_{44}^{e_6^{(1)}}S_{44}^{e_6^{(1)}}
|_{h\rightarrow H}
=1^2~3^4~5^6|_{h\rightarrow H}=\{1\}^2\{11\}^2\{3\}^4\{9\}^4\{5\}^6\{7\}^6
|_{h\rightarrow H}\nonumber\\   
&&=\{1\}\{h-1\}\{h/3-3\}\{2h/3+3\}\{3\}\{h-3\}\nonumber\\
&&~\times\{h/3-1\}^2\{2h/3+1\}^2\{2h/3-5\}
\{h/3+5\}\{5\}\nonumber\\
&&~\times\{h-5\}\{h/3+1\}^3\{2h/3-1\}^3
\{2h/3-3\}^2\{h/3+3\}^2|_{h\rightarrow H}\nonumber\\
&&=\{1\}_0\{H-1\}_0\{H/3-3\}_0\{2H/3+3\}_0\{3\}_0\{H-3\}_0
\{H/3-1\}_0^2\nonumber\\
&&~\times\{2H/3+1\}_0^2\{2H/3-5\}_0\{H/3+5\}_0\{5\}_0\{H-5\}_0\nonumber\\
&&~\times\{H/3+1\}_0^3\{2H/3-1\}_0^3
\{2H/3-3\}_0^2\{H/3+3\}_0^2|_{h\rightarrow H}\nonumber\\
&&=[1]_0[3]_0[5]_0[H/3-3]_0[H/3-1]_0^2[H/3+1]_0^3[H/3+3]_0^2[H/3+5]_0
\Enar

For showing the \rref{nfora}--\rref{nfor} the calculations 
are done in following fashion.
\Bear
&S_{14}^{f_4^{(1)}}&=S_{14}^{e_6^{(1)}}|_{h\rightarrow H}
=3~5|_{h\rightarrow H}=\{3\}\{9\}\{5\}\{7\}|_{h\rightarrow H}\nonumber\\
&&={\frac{(2)(4)}{(2+B)(4-B)}}{\frac{(8)(10)}{(8+B)(10-B)}}
{\frac{(4)(6)}{(4+B)(6-B)}}{\frac{(6)(8)}{(6+B)(8-B)}}
{\Big |}_{h\rightarrow H}\nonumber\\
&&={\frac{(2)(h/3)}{(h/3-2+B)(4-B)}}{\frac{(2h/3)(h-2)}
{(h-4+B)(2h/3+2-B)}}\nonumber\\
&&~\times{\frac{(h/3)(2h/3-2)}{(2h/3-4+B)(h/3+2-B)}}
{\frac{(h/3+2)(2h/3)}
{(2h/3-2+B)(h/3+4-B)}}{\Big |}_{h\rightarrow H}\nonumber\\
&&=\hcd{2}\hcd{2h/3}\hcd{h/3}\hcd{h/3+2}|_{h\rightarrow H}
=\hcd{2}_0\hcd{2H/3}_0\hcd{H/3}_0\hcd{H/3+2}_0\nonumber\\
&&=\{H/6+1\}_{1/2}\{5H/6-1\}_{1/2}\{H/2-1\}_{1/2}
\{H/2+1\}_{1/2}\nonumber\\
&&=[H/6+1]_{1/2}[H/2+1]_{1/2}\\
&S_{22}^{f_4^{(1)}}&=S_{22}^{e_6^{(1)}}|_{h\rightarrow H}
=1~5|_{h\rightarrow H}=\{1\}\{11\}\{5\}\{7\}|_{h\rightarrow H}\nonumber\\
&&={\frac{(0)(2)}{(B)(2-B)}}{\frac{(10)(12)}{(10+B)(12-B)}}
{\frac{(4)(6)}{(4+B)(6-B)}}{\frac{(6)(8)}{(6+B)(8-B)}}
{\Big |}_{h\rightarrow H}\nonumber\\
&&={\frac{(0)(h/3-2)}{(h/3-4+B)(2-B)}}{\frac{(2h/3+2)(h)}
{(h-2+B)(2h/3+4-B)}}\nonumber\\
&&~\times{\frac{(h/3)(2h/3-2)}{(2h/3-4+B)(h/3+2-B)}}
{\frac{(h/3+2)(2h/3)}
{(2h/3-2+B)(h/3+4-B)}}{\Big |}_{h\rightarrow H}\nonumber\\
&&=\hcd{0}\hcd{2h/3+2}\hcd{h/3}\hcd{h/3+2}|_{h\rightarrow H}\nonumber\\
&&=\hcd{0}_0\hcd{2H/3+2}_0\hcd{H/3}_0\hcd{H/3+2}_0\nonumber\\
&&=\{H/6-1\}_{1/2}\{5H/6+1\}_{1/2}\{H/2-1\}_{1/2}
\{H/2+1\}_{1/2}\nonumber\\
&&=[H/6-1]_{1/2}[H/2+1]_{1/2}\\
&S_{23}^{f_4^{(1)}}&=S_{23}^{e_6^{(1)}}|_{h\rightarrow H}
=3~5|_{h\rightarrow H}=\{3\}\{9\}\{5\}\{7\}|_{h\rightarrow H}\nonumber\\
&&={\frac{(2)(4)}{(2+B)(4-B)}}{\frac{(8)(10)}{(8+B)(10-B)}}
{\frac{(4)(6)}{(4+B)(6-B)}}{\frac{(6)(8)}{(6+B)(8-B)}}
{\Big |}_{h\rightarrow H}\nonumber\\
&&=\hcd{h/6}\hcd{h/2+2}\hcd{h/6+2}\hcd{h/2}|_{h\rightarrow H}\nonumber\\
&&=\hcd{H/6}_0\hcd{H/2+2}_0\hcd{H/6+2}_0\hcd{H/2}_0\nonumber\\
&&=\{H/3-1\}_{1/2}\{2H/3+1\}_{1/2}\{H/3+1\}_{1/2}
\{2H/3-1\}_{1/2}\nonumber\\
&&=[H/3-1]_{1/2}[H/3+1]_{1/2}\\
&S_{24}^{f_4^{(1)}}&=S_{24}^{e_6^{(1)}}|_{h\rightarrow H}
=2~4~6|_{h\rightarrow H}=\{2\}\{10\}\{4\}\{8\}\{6\}\{6\}
|_{h\rightarrow H}\nonumber\\
&&=\hcd{h/6-1}\hcd{h/2+3}\hcd{h/6+1}\hcd{h/2+1}\hcd{h/6+3}\hcd{h/2-1}
|_{h\rightarrow H}\nonumber\\
&&=\hcd{H/6-1}_0\hcd{H/2+3}_0\hcd{H/6+1}_0\hcd{H/2+1}_0\hcd{H/6+3}_0
\hcd{H/2-1}_0\nonumber\\
&&=\{H/3-2\}_{1/2}\{2H/3+2\}_{1/2}\{H/3\}_{1/2}\{2H/3\}_{1/2}
\{H/3+2\}_{1/2}\{2H/3-2\}_{1/2}\nonumber\\
&&=[H/3-2]_{1/2}[H/3]_{1/2}[H/3+2]_{1/2}\\
&S_{34}^{f_4^{(1)}}&=S_{34}^{e_6^{(1)}}|_{h\rightarrow H}
=2~4^2~6|_{h\rightarrow H}=\{2\}\{10\}\{4\}\{8\}\{4\}\{8\}
\{6\}\{6\}|_{h\rightarrow H}\nonumber\\
&&=\hcd{1}\hcd{2h/3+1}\hcd{3}\hcd{2h/3-1}\hcd{h/3-1}\hcd{h/3+3}
\hcd{h/3+1}^2|_{h\rightarrow H}\nonumber\\
&&=\hcd{1}_0\hcd{2H/3+1}_0\hcd{3}_0\hcd{2H/3-1}_0\hcd{H/3-1}_0\hcd{H/3+3}
_0\hcd{H/3+1}_0^2\nonumber\\
&&=\{H/6\}_{1/2}\{5H/6\}_{1/2}\{H/6+2\}_{1/2}\{5H/6-2\}_{1/2}\nonumber\\
&&~\times\{H/2-2\}_{1/2}\{H/2+2\}_{1/2}\{H/2\}_{1/2}^2\nonumber\\
&&=[H/6]_{1/2}[H/6+2]_{1/2}[H/2+2]_{1/2}[H/2]_{1/2}\\
&S_{44}^{f_4^{(1)}}&=S_{44}^{e_6^{(1)}}|_{h\rightarrow H}
=1~3^2~5^3|_{h\rightarrow H}=\{1\}\{11\}\{3\}^2\{9\}^2\{5\}^3
\{7\}^3|_{h\rightarrow H}\nonumber\\
&&=\hcd{0}\hcd{2h/3+2}\hcd{2}\hcd{h/3-2}\hcd{2h/3}\hcd{h/3+4}
\hcd{4}\nonumber\\
&&~\times\hcd{h/3}^2\hcd{2h/3-2}\hcd{h/3+2}^2|_{h\rightarrow H}\nonumber\\
&&=\hcd{0}_0\hcd{2H/3+2}_0\hcd{2}_0\hcd{H/3-2}_0\hcd{2H/3}_0\hcd{H/3+4}_0
\nonumber\\
&&~\times\hcd{4}_0\hcd{H/3}^2_0\hcd{2H/3-2}_0\hcd{H/3+2}_0^2\nonumber\\
&&=\{H/6-1\}_{1/2}\{5H/6+1\}_{1/2}\{H/6+1\}_{1/2}\{H/2-3\}_{1/2}
\{5H/6-1\}_{1/2}\nonumber\\
&&\times\{H/2+3\}_{1/2}\{H/6+3\}_{1/2}
\{H/2-1\}_{1/2}^2\{5H/6-3\}_{1/2}\{H/2+1\}_{1/2}^2\nonumber\\
&&=[H/6-1]_{1/2}[H/6+1]_{1/2}[H/2-3]_{1/2}[H/6+3]_{1/2}[H/2+1]_{1/2}^2
\nonumber\\
&&=[H/6-1]_{1/2}[H/6+1]_{1/2}[H/3]_0^{'}[H/2+1]_{1/2}^2
\Enar 

\section*{Appendix B: $e_7^{(1)}\rightarrow e_6^{(2)}$}
\setcounter{equation}{0}
\renewcommand{\theequation}{B.\arabic{equation}}
\Bear
&S_{12}^{e_6^{(2)}}&=S_{24}^{e_7^{(1)}}|_{h\rightarrow H}
=5~7|_{h\rightarrow H}=\{5\}\{13\}\{7\}\{11\}|_{h\rightarrow H}\nonumber\\
&&=\{h/6+2\}\{5h/6-2\}\{h/2-2\}\{h/2+2\}|_{h\rightarrow H}\nonumber\\
&&=\{H/6+2\}_0\{5H/6-2\}_0\{H/2-2\}_0\{H/2+2\}_0\nonumber\\
&&=[H/6+2]_0[H/2+2]_0\\
&S_{13}^{e_6^{(2)}}&=S_{25}^{e_7^{(1)}}|_{h\rightarrow H}
=2~6~8|_{h\rightarrow H}=\{2\}\{16\}\{6\}\{12\}\{8\}\{10\}
|_{h\rightarrow H}\nonumber\\
&&=\{2\}\{h-2\}\{h/3\}\{2h/3\}\{h/3+2\}\{2h/3-2\}|_{h\rightarrow H}
\nonumber\\
&&=\{2\}_0\{H-2\}_0\{H/3\}_0\{2H/3\}_0\{H/3+2\}_0\{2H/3-2\}_0\nonumber\\
&&=[2]_0[H/3]_0[H/3+2]_0\\
&S_{14}^{e_6^{(2)}}&=S_{27}^{e_7^{(1)}}|_{h\rightarrow H}
=3~5~7~9|_{h\rightarrow H}=\{3\}\{15\}\{5\}\{13\}\{7\}\{11\}\{9\}\{9\}
|_{h\rightarrow H}\nonumber\\
&&=\{3\}\{h-3\}\{h/3-1\}\{2h/3+1\}\{h/3+1\}\nonumber\\
&&~\times\{2h/3-1\}\{h/3+3\}\{2h/3-3\}
|_{h\rightarrow H}\nonumber\\
&&=\{3\}_0\{H-3\}_0\{H/3-1\}_0\{2H/3+1\}_0\{H/3+1\}_0\nonumber\\
&&~\times\{2H/3-1\}_0\{H/3+3\}_0\{2H/3-3\}_0\nonumber\\
&&=[3]_0[H/3-1]_0[H/3+1]_0[H/3+3]_0\\
&S_{22}^{e_6^{(2)}}&=S_{44}^{e_7^{(1)}}|_{h\rightarrow H}
=1~3~7~9|_{h\rightarrow H}=\{1\}\{17\}\{3\}\{15\}\{7\}\{11\}\{9\}\{9\}
|_{h\rightarrow H}\nonumber\\
&&=\{1\}\{h-1\}\{h/3-3\}\{2h/3+3\}\{h/3+1\}\nonumber\\
&&~\times\{2h/3-1\}\{h/3+3\}\{2h/3-3\}
|_{h\rightarrow H}\nonumber\\
&&=\{1\}_0\{H-1\}_0\{H/3-3\}_0\{2H/3+3\}_0\{H/3+1\}_0\nonumber\\
&&~\times\{2H/3-1\}_0\{H/3+3\}_0\{2H/3-3\}_0\nonumber\\
&&=[1]_0[H/3-3]_0[H/3+1]_0[H/3+3]_0\\
&S_{23}^{e_6^{(2)}}&=S_{45}^{e_7^{(1)}}|_{h\rightarrow H}
=4~6^2~8|_{h\rightarrow H}=\{4\}\{14\}\{6\}\{12\}\{6\}\{12\}\{8\}\{10\}
|_{h\rightarrow H}\nonumber\\
&&=\{h/6+1\}\{5h/6-1\}\{h/6+3\}\{5h/6-3\}\nonumber\\
&&~\times\{h/2-3\}\{h/2+3\}\{h/2-1\}\{h/2+1\}
|_{h\rightarrow H}\nonumber\\
&&=\{H/6+1\}_0\{5H/6-1\}_0\{H/6+3\}_0\{5H/6-3\}_0\nonumber\\
&&~\times\{H/2-3\}_0\{H/2+3\}_0\{H/2-1\}_0\{H/2+1\}_0
\nonumber\\
&&=[H/6+1]_0[H/6+3]_0[H/2+1]_0[H/2+3]_0\\
&S_{24}^{e_6^{(2)}}&=S_{47}^{e_7^{(1)}}|_{h\rightarrow H}
=3~5^2~7^2~9|_{h\rightarrow H}=\{3\}\{15\}\{5\}^2\{13\}^2\{7\}^2
\{11\}^2\{9\}\{9\}|_{h\rightarrow H}\nonumber\\
&&=\{h/6\}\{5h/6\}\{h/6+2\}\{5h/6-2\}\{h/2-4\}\{h/2+4\}\nonumber\\
&&~\times\{h/6+4\}\{5h/6-4\}\{h/2-2\}\{h/2+2\}\{h/2\}\{h/2\}
|_{h\rightarrow H}\nonumber\\
&&=\{H/6\}_0\{5H/6\}_0\{H/6+2\}_0\{5H/6-2\}_0
\{H/2-4\}_0\{H/2+4\}_0\nonumber\\
&&~\times\{H/6+4\}_0\{5H/6-4\}_0\{H/2-2\}_0\{H/2+2\}_0
\{H/2\}_0\{H/2\}_0\nonumber\\
&&=[H/6]_0[H/6+2]_0[H/2+4]_0[H/6+4]_0[H/2+2]_0[H/2]_0\\
&S_{33}^{e_6^{(2)}}&=S_{55}^{e_7^{(1)}}|_{h\rightarrow H}
=1~3~5~7^2~9|_{h\rightarrow H}=\{1\}\{17\}\{3\}\{15\}\{5\}
\{13\}\{7\}^2\{11\}^2\{9\}\{9\}|_{h\rightarrow H}\nonumber\\
&&=\{1\}\{h-1\}\{3\}\{h-3\}\{h/3-1\}\{2h/3+1\}\nonumber\\
&&~\times\{h/3+1\}^2\{2h/3-1\}^2\{h/3+3\}\{2h/3-3\}
|_{h\rightarrow H}\nonumber\\
&&=\{1\}_0\{H-1\}_0\{3\}_0\{H-3\}_0\{H/3-1\}_0\{2H/3+1\}_0\nonumber\\
&&~\times\{H/3+1\}_0^2\{2H/3-1\}_0^2\{H/3+3\}_0\{2H/3-3\}_0\nonumber\\
&&=[1]_0[3]_0[H/3-1]_0[H/3+1]_0^2[H/3+3]_0\\
&S_{34}^{e_6^{(2)}}&=S_{57}^{e_7^{(1)}}|_{h\rightarrow H}
=2~4^2~6^2~8^3|_{h\rightarrow H}=\{2\}\{16\}\{4\}^2\{14\}^2
\{6\}^2\{12\}^2\{8\}^3\{10\}^3|_{h\rightarrow H}\nonumber\\
&&=\{2\}\{h-2\}\{4\}\{h-4\}\{h/3-2\}\{2h/3+2\}\nonumber\\
&&~\times\{h/3\}^2\{2h/3\}^2\{h/3+2\}^2\{2h/3-2\}^2\{2h/3-4\}\{h/3+4\}
|_{h\rightarrow H}\nonumber\\
&&=\{2\}_0\{H-2\}_0\{4\}_0\{H-4\}_0\{H/3-2\}_0\{2H/3+2\}_0\nonumber\\
&&~\times\{H/3\}_0^2\{2H/3\}_0^2\{H/3+2\}_0^2\{2H/3-2\}_0^2
\{2H/3-4\}_0\{H/3+4\}_0
\nonumber\\
&&=[H/6]_0[H/6+2]_0[H/2+4]_0[H/6+4]_0[H/2+2]_0[H/2]_0\\
&S_{44}^{e_6^{(2)}}&=S_{77}^{e_7^{(1)}}|_{h\rightarrow H}
=1~3^2~5^3~7^4~9^2|_{h\rightarrow H}\nonumber\\
&&=\{1\}\{17\}\{3\}^2\{15\}^2\{5\}^3\{13\}^3\{7\}^4\{11\}^4
\{9\}^2\{9\}^2|_{h\rightarrow H}\nonumber\\
&&=\{1\}\{h-1\}\{3\}\{h-3\}\{h/3-3\}\{2h/3+3\}\{5\}\nonumber\\
&&~\times\{h-5\}\{h/3-1\}^2\{2h/3+1\}^2\{h/3+1\}^3\{2h/3-1\}^3\nonumber\\
&&~\times\{2h/3-5\}\{h/3+5\}\{h/3+3\}^2\{2h/3-3\}^2
|_{h\rightarrow H}\nonumber\\
&&=\{1\}_0\{H-1\}_0\{3\}_0\{H-3\}_0\{H/3-3\}_0
\{2H/3+3\}_0\{5\}_0\nonumber\\
&&~\times\{H-5\}_0\{H/3-1\}_0^2\{2H/3+1\}_0^2
\{H/3+1\}_0^3\{2H/3-1\}_0^3\nonumber\\
&&~\times\{2H/3-5\}_0\{H/3+5\}_0\{H/3+3\}_0^2\{2H/3-3\}_0^2
\nonumber\\
&&=[1]_0[3]_0[H/3-3]_0[5]_0[H/3-1]_0^2[H/3+1]_0^3[H/3+5]_0[H/3+3]_0^2
\Enar


\end{document}